\documentclass{article}
\usepackage{natbib}
\usepackage{arxiv}

\usepackage[utf8]{inputenc} 
\usepackage[T1]{fontenc}    
\usepackage{hyperref}       
\usepackage{booktabs}       
\usepackage{ulem}
\usepackage{tabularray}
\usepackage{multirow}
\usepackage{amsfonts}       
\usepackage{nicefrac}       
\usepackage{microtype}      
\usepackage{lipsum}
\usepackage{fancyhdr}       

\usepackage{times}  
\usepackage{helvet}  
\usepackage{courier}  
\usepackage{amsthm}
\usepackage{algorithm}
\usepackage{algorithmic}
\usepackage[switch]{lineno}
\usepackage{lipsum}
\usepackage{makecell}
\usepackage{subcaption}
\usepackage{graphicx}
\usepackage{xurl}

\title{Predicting Internet Connectivity in Schools: A Feasibility Study Leveraging Multi-modal Data and Location Encoders in Low-Resource Settings}

\author{
  Kelsey Doerksen \\
  UNICEF \\
  University of Oxford \\
  \texttt{kelsey.doerksen@cs.ox.ac.uk} \\
   \And
  Casper Fibaek \\
  European Space Agency $\phi$-lab \\
  \texttt{casper.fibaek@esa.int} \\
  \And
  Rochelle Schneider \\
  European Space Agency $\phi$-lab  \\
  \texttt{rochelle.schneider@esa.int} \\
  \And
  Do-Hyung Kim \\
  UNICEF \\
  \texttt{dokim@unicef.org} \\
  \And
  Isabelle Tingzon \\
  UNICEF \\
  \texttt{itingzon@unicef.org} \\
}

\begin{document}
\maketitle

\begin{abstract}
Internet connectivity in schools is critical to provide students with the digital literary skills necessary to compete in modern economies. In order for governments to effectively implement digital infrastructure development in schools, accurate internet connectivity information is required. However, traditional survey-based methods can exceed the financial and capacity limits of governments. Open-source Earth Observation (EO) datasets have unlocked our ability to observe and understand socio-economic conditions on Earth from space, and in combination with Machine Learning (ML), can provide the tools to circumvent costly ground-based survey methods to support infrastructure development. In this paper, we present our work on school internet connectivity prediction using EO and ML. We detail the creation of our multi-modal, freely-available satellite imagery and survey information dataset, leverage the latest geographically-aware location encoders, and introduce the first results of using the new European Space Agency $\phi$-lab geographically-aware foundational model to predict internet connectivity in Botswana and Rwanda. We find that ML with EO and ground-based auxiliary data yields the best performance in both countries, for accuracy, F1 score, and False Positive rates, and highlight the challenges of internet connectivity prediction from space with a case study in Kigali, Rwanda. Our work showcases a practical approach to support data-driven digital infrastructure development in low-resource settings, leveraging freely available information, and provide cleaned and labelled datasets for future studies to the community through a unique collaboration between UNICEF and the European Space Agency $\phi$-lab.
\end{abstract}

\section{Introduction}
In December 2020, the United Nations Children’s Fund (UNICEF) and the International Telecommunication Union (ITU) reported that two-thirds (or 1.3 billion) of the world's school-age children do not have an internet connection in their homes \citep{unicefitu2024}. To combat this, the \textit{Giga} initiative was created, aiming to connect every school to the internet and every young person to information, opportunity and choice \citep{Giga}. Worldwide, schools provide critical online learning infrastructure to communities, and the digital divide between lower-income countries and more developed nations exacerbates already existing inequalities, causing children to fall even further behind. Our work targets the United Nations' Sustainable Development Goal (SDG) 4 of Quality Education: \textit{Ensuring inclusive and equitable quality education and promote lifelong learning opportunities for all}. The digital divide refers to the disparity between countries, regions, and people and their access to digital services \citep{digitaldivide}. In 2023, 60\% of the African population is unconnected to the global digital society \citep{worldinterent}, and in the United Nations-designated least developed countries, less than 30\% of the population uses the Internet, according to a 2021 estimate \citep{ITU}. Digital access, information and communication are key to a country's development, and this gross disparity is considered a global challenge \citep{Bon2024}. Traditionally, policymakers rely on prohibitively costly and timely surveys to capture the data required for a clear understanding of a country's digital infrastructure. Earth Observation can provide governments and stakeholders an affordable, objective alternative to support infrastructure and sustainable development programs. EO is the process of gathering information about the physical, chemical, and biological systems of the planet via remote-sensing \citep{esaEO}. Open-source, satellite-based measurements have traditionally been under-utilized in low-income contexts, where such data can provide great benefits ~\citep{metadisc,haack2016improving}. EO data is particularly powerful in data-scarce developing countries, complementing other sources of data such as censuses and ground teams \citep{esa2030}. Large volumes of EO data are now easily accessible and can provide unparalleled information about Earth systems and society, creating the opportunity to leverage Machine Learning to reveal patterns and insights across multiple domains \citep{9553464}. It is not, however, practical to assume that high-resolution, pre-processed data can be seamlessly obtained and utilized. A key motivation and contribution of our work is the utilization of \textit{free}, globally-available data sources, for which we have created a labelled dataset for future studies for the broader community.

Within Artificial Intelligence the (AI) for EO communities, there has been an increased focus on developing geographically aware AI models. These models come in various forms, but shared amongst them is that they, given images or locations, provide embeddings that summarize the characteristics of a given area. These embeddings offer convenient usage in diverse downstream tasks \citep{satclip}. The embeddings are directly helpful for classification and clustering, and the models can be fine-tuned with new prediction heads to provide end-to-end workflows. These vector embeddings, referred to as \textit{feature embeddings} throughout this work, have performed well in various predictive tasks, ranging from multi-class classification to image recognition and generalizing geographically. By pre-training models with global data and location information, the aim is to make the models robust towards variance in geography, providing improved generalizability across geographical domains. Geographic generalization is a critical component of the adoption of these AI methodologies in developing countries, where often economic disparities between developing nations and the Global North result in critical infrastructure gaps in training data availability, resulting in models trained on datasets well representing Europe and North America, and failing in contexts within the Global South. 

We present work on ML-powered school internet connectivity prediction leveraging EO data, to support data-driven infrastructure development in digitally-underserved communities. Contributing to the AI for EO community, we show our results of hand-crafted engineered feature spaces using a combination of fully open-source EO and survey data compared to and in combination with the use of geographical-aware location encoder-extracted feature embeddings as inputs to our ML classifiers, and introduce a new geographical AI Foundation model. We demonstrate our model's performance in two pilot countries, Botswana and Rwanda. To provide digestible and interpretable outputs for stakeholders, we highlight feature importance metrics for connectivity prediction and provide intuitive model outputs in the form of maps to easily visualize where and who needs support. In the context of this work, we refer to connectivity as the usage of Internet by an entity (school), with no boundaries on performance (i.e. upload or download speeds).

\section{Related Work}
\textbf{Machine Learning for Social Good with Earth Observation:}
Over the past decade, the theme of AI for Social Good has grown rapidly, whereby researchers aim at developing AI methods and tools to address problems at the societal level and improve the well-being of the society \citep{shi2020artificial}. Inferring socio-economic indicators with ML and space-based datasets has been investigated, with numerous works focused on poverty mapping ~\citep{phil-pov,PetterssonKOJD23,pubsatafrica}. Mapping infrastructure with EO and ML is another popular area of research, with studies focused on identifying informal settlements, slums, and deprived areas in ~\citep{fallatah2020object,wurm2019semantic,wang2019role,rs14133072}, but there has been relatively little attention in the domain of connectivity mapping. The World Bank has created an openly-available composite map of the global power system with ML \citep{predelec}, and there has been published work on policy evaluation in data-sparse environments to assess the livelihood impact of electricity access in Uganda using satellite imagery and ML \citep{eleclive}, but there have been no studies focused on internet connectivity mapping. Our work builds on the previous work of socio-economic infrastructure mapping to provide an analysis of the feasability and limitations of ML for estimating internet connectivity in schools with freely-available satellite data, and tests the capability of global, general-purpose location embeddings from CLIP-based models to support our use case.

\noindent\textbf{Geographically-Aware Neural Networks:}
Incorporating spatial context within AI models using satellite imagery can improve model generalization and overall performance, and growing interest in developing geographically-aware embeddings from geospatial datasets has been explored in the development of the Satellite Contrastive Location-Image Pretraining (SatCLIP), GeoCLIP, Contrastive Spatial Pre-Training (CSP), and GPS2Vec models ~\citep{satclip,geoclip,csp,gps2vec}. SatCLIP, GeoCLIP and CSP all utilize the concept of Contrastive Language-Image Pretraining (CLIP), whereby models are trained on a variety of (image, text) pairs \citep{radford2021learning}. CLIP is an efficient method of image representation learning from natural language supervision. This concept is extended to incorporating geographic context with EO data in these models, whereby instead of training text and image encoders, a geographically-aware location encoder is trained to learn implicit representations of locations from satellite imagery. SatCLIP in particular has previously showcased its ability to perform well on a variety of downstream tasks utilizing open-source satellite imagery, including Median Income, California Housing Prices and logged Population Density ~\citep{10.1145/3394486.3403101,RePEc:eee:stapro:v:33:y:1997:i:3:p:291-297,generalmlsat}. Our work leverages the SatCLIP, GeoCLIP and CSP location encoders for a comparative analysis of suitability for internet connectivity prediction in schools.

\section{Data}
\subsection{Engineered Features}
\textbf{Satellite Imagery Features:} We derived a rich set of tabularized features from publicly available mid-resolution satellite data from Google Earth Engine (GEE). Taking each location with connectivity information in our dataset as the center point, we extract a 1,000m radius extent of satellite data from GEE including MODIS landcover, VIIRS Nightlight, Global Human Modification, Gridded Population of the World, and Global Human Settlement Layer data products using the airPy\footnote{\tiny{\url{https:/github.com/kelsdoerksen/airPy}}} data processing package ~\cite{modislc,nightlight,ghm,popdata,ghsl}, which represents the land use and extent of urbanization surrounding the schools. We relate our work to poverty mapping, with the assumption that regions suffering from greater levels of poverty are less developed and therefore less likely to have internet access. Land-use data has been used in previous poverty mapping studies by \citep{TIAN20227} which highlighted the relationship between cropland ratio and poverty mapping, and nighttime light intensity and building distribution data has been used by \citep{PUTRI2023100889} with ML for poverty mapping in East Java, Indonesia, providing support for our data selection choices. Table \ref{table:school-conn-features} summarizes the engineered feature space. We selected the geospatial buffer extent for feature creation by first performing a comparative analysis of ML classifier connectivity prediction performance using only geospatial features for 300m, 500m, 750m, 1,000m, and 5,000m buffers extents and found that 1,000m yielded the best results for test set F1 score and accuracy. Further details on feature generation, datasets, and buffer selection are found in Supplementary Materials.

\begin{table*}[ht!]
\small
\centering
\caption{School connectivity geospatial feature space.} 
\label{table:school-conn-features}
\centering
{
\begin{tabular}{lll}
\toprule
\textbf{Source} & \textbf{Resolution} & \textbf{Metrics} \\
\midrule
MODIS Land cover & 500m & \makecell[l]{\% coverage per class, mode, variance} \\  
Gridded Population of the World  & 927.67m & \makecell[l]{mean, variance, max, min} \\
VIIRS Nightlight  & 463.83m & \makecell[l]{max, min, mean, variance} \\
Global Human Settlement Layer & 10m & \% coverage per class, mode, variance \\
Global Human Modification & 1000m & mode, variance, mean, max, min \\
World Bank Electrical Power Grid & - & distance from school to nearest transmission line \\
Ookla Speedtest & - & \makecell[l]{average download, upload and latency for mobile and fixed tests, \\ no. of mobile and fixed tests per tile, \\ distance of nearest tile to school, no. of unique mobile and fixed devices} \\
\bottomrule
\end{tabular}}
\end{table*}

\noindent\textbf{Ground-based Infrastructure Features:} To integrate power grid information, the distance from each school to the nearest transmission line is calculated and added as a feature using electrical power grid information from the World Bank Group \citep{predelec}. To incorporate mobile and fixed device speed test data, we calculate the distance of each school to the nearest mobile and fixed broadband Ookla tile and include relevant information about download and upload speeds from devices recorded in that tile, with data provided by the Ookla for good initiative from October 2023 \citep{ookla}.

\noindent\textbf{Data Pre-processing:} We perform min-max scaling of the feature space to improve performance as implemented in \textit{scikit-learn}. Using the pearson standard correlation coefficient, we compute the pairwise correlation of feature columns and remove those with a score of greater than 0.9.

\begin{figure*}[!htbp]
    \centering
    {\includegraphics[scale=0.38]{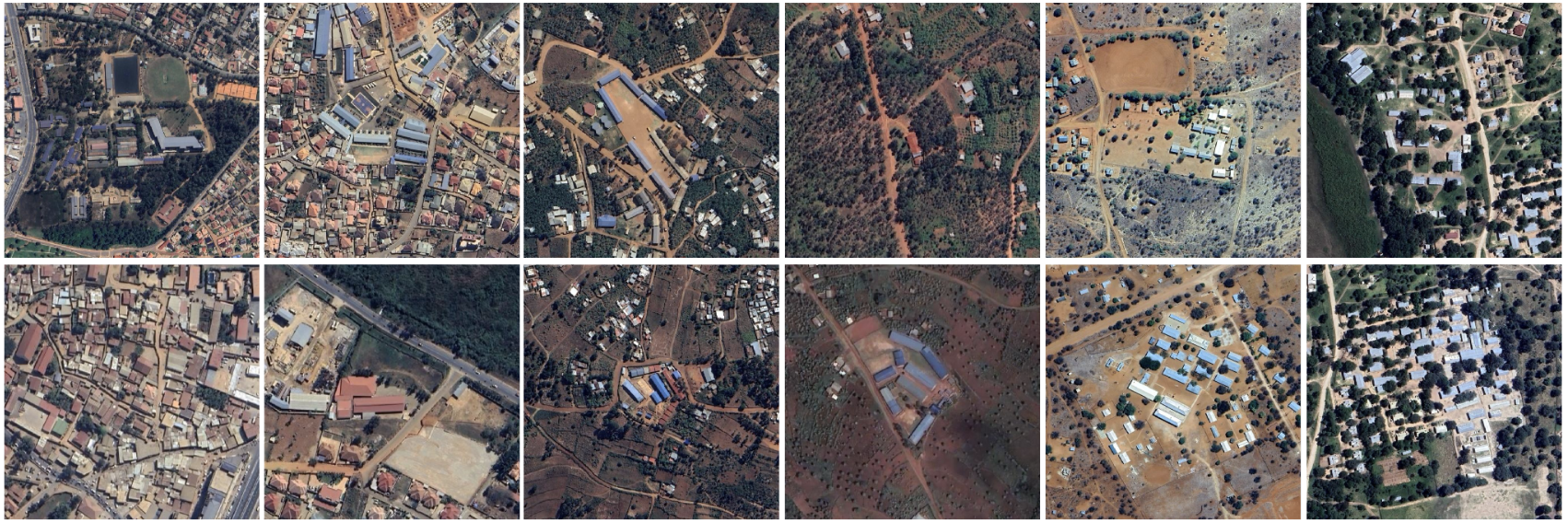}}
    \caption{From left to right, Top: 3 connected schools in Rwanda, 2 connected schools in Botswana. Bottom: 3 unconnected schools in Rwanda, 2 unconnected schools in Botswana.
    \label{fig:connvsnon}}
\end{figure*}
\noindent\textbf{Labels:} Location and connectivity status information per institution is provided from government partners through Project Connect, which provides the longitude, latitude and binary connectivity label for each school in the dataset \citep{projconnect}. Project Connect aims to map real-time school connectivity and provide this critical information to governments and internet service providers (ISPs) globally. Table \ref{table:school-conn-class-dist} summarizes the class balance of connected/unconnected schools in our study countries Botswana and Rwanda. Figure \ref{fig:connvsnon} shows 6 visual examples of connected (top row) and unconnected (bottom row) schools.

\begin{table}
\centering
\small
\caption{The class distribution (Yes = Connected to internet, No = Not connected to internet) across the training, validation and test sets of Botswana (\textbf{BWA}) and Rwanda (\textbf{RWA}).}
\label{table:school-conn-class-dist}
{%
\begin{tblr}{
  column{even} = {c},
  column{3} = {c},
  column{5} = {c},
  column{7} = {c},
  cell{1}{2} = {c=2}{},
  cell{1}{4} = {c=2}{},
  cell{1}{6} = {c=2}{},
  hline{1,5} = {-}{},
  hline{3} = {2-7}{},
}
             & \textbf{Training (70\%)} &             & \textbf{Validation (15\%)} &             & \textbf{Test (15\%)} &             \\
             & \textbf{Yes}                 & \textbf{No} & \textbf{Yes}                   & \textbf{No} & \textbf{Yes}             & \textbf{No} \\
\textbf{BWA} & 327                          & 307         & 74                             & 62          & 75                       & 62          \\
\textbf{RWA} & 1317                         & 1025        & 295                            & 207         & 276                      & 235         
\end{tblr}
}
\end{table}

\subsection{Satellite Imagery to Feature Embeddings}
\begin{figure*}[!htbp]
    \centering
    {\setlength\intextsep{0pt}
    {\includegraphics[scale=0.50]{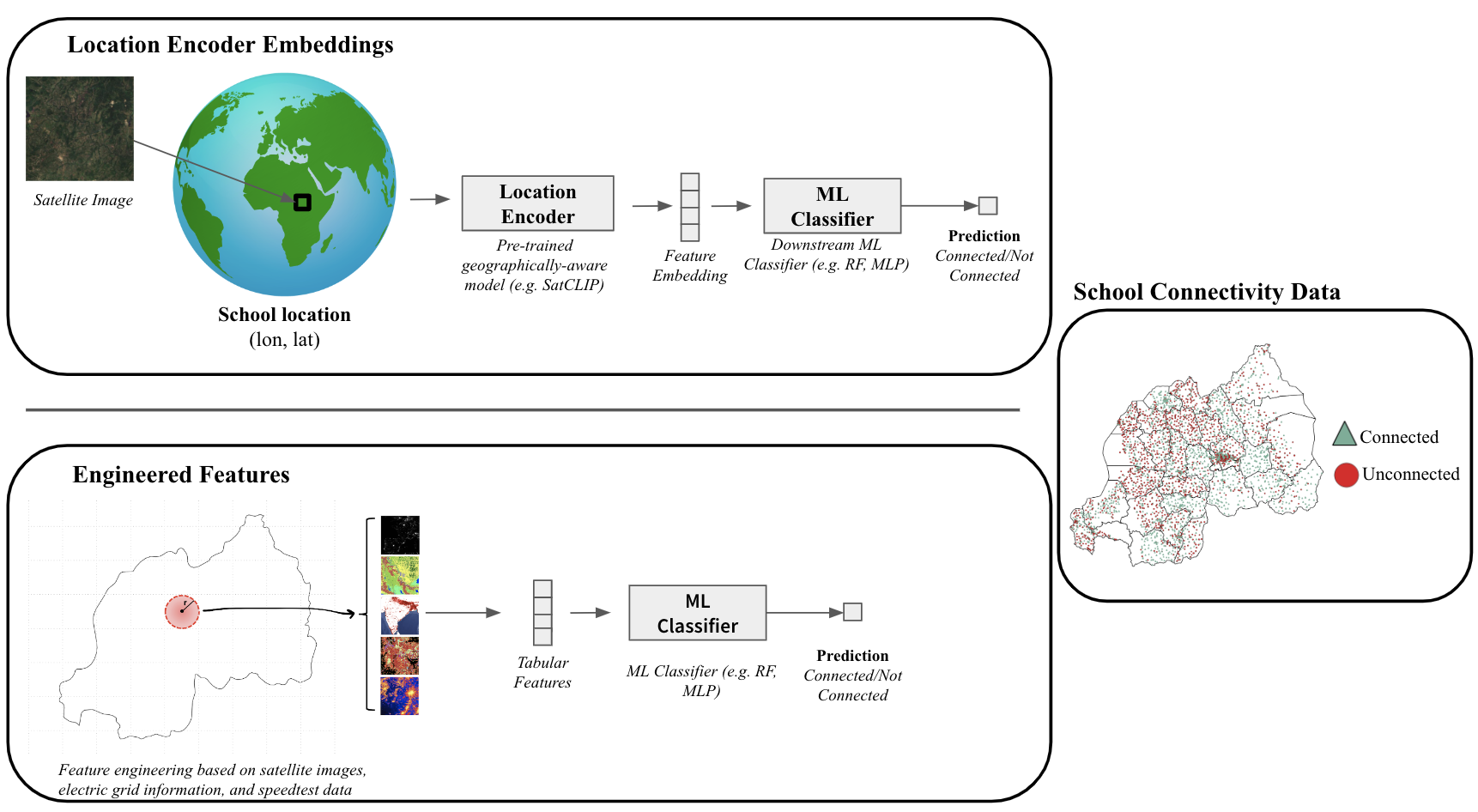}}
    \caption{Methodological workflow for connectivity prediction using engineered features and location encoder embeddings. \textbf{Top}: Per latitude, longitude point representing a school location with known internet connectivity label, generate feature embedding for the corresponding satellite image geo-referenced to the point, per location encoder. \textbf{Bottom}: Per latitude, longitude point representing a school location with known internet connectivity label and a user-specified buffer, extract the satellite imagery within the buffer boundaries and calculate summary statistics of the geospatial information to create tabular features. \label{fig:dataset}}}
\end{figure*}

\noindent\textbf{Problem Formulation: School Connectivity Prediction:} We model connectivity prediction as a binary classification task, wherein we classify each sample (school) represented by a feature vector of engineered features, location encoder model embeddings, or a combination based on its internet connectivity status. We have a dataset $D$ which contains $(x_i, y_i)_i$ where $x_i$ in $X$ is a vector of features representing a combination of information extracted from satellite imagery (e.g. land cover, nightlight), survey and ground-based measurements (e.g. Ookla Speedtest, global transmission line network) and/or embeddings from location encoder models about a school $I$ and $y_i$ in $Y$ is the internet connectivity status (i.e. 1 connected, 0 not connected) we aim to predict. Figure \ref{fig:dataset} depicts a overview of the methodological workflow of feature generation and connectivity classification.

We explore the use of vector embeddings extracted from the location encoders of three pre-existing models as inputs into our ML classifiers for connectivity prediction, namely SatCLIP, GeoCLIP and CSP ~\citep{satclip,geoclip,csp}, and our newly developed PhilEO Very High Resolution (VHR) Pre-cursor model with embedding sizes of 256, 512, 256, and 1024, respectively. Vector embeddings extracted from images can be thought of as extremely small summaries of those images which claim to capture the ground conditions including population density, housing infrastructure, sun exposure, and more \citep{satclip}. Each location encoder used in our study is trained with a different dataset to explore performance differences between imagery sources. CSP was selected for its pre-training on the Functional Map of the World (FMoW) datatset, which was developed to predict the functional purpose of buildings and land use from satellite images and complementary metadata features \citep{2017arXiv171107846C}. Categories in FMoW are grouped according to their purpose, and it is proposed that there is potential to delineating connectivity status from buildings (in our case schools), by leveraging the embeddings from a pre-trained model on this dataset. The CSP model is also available in its pre-trained form on the iNaturalist dataset \citep{inaturalist}, however this was not pursued as the dataset is not appropriate for our use case. SatCLIP was selected because of its superior performance to other location encoder techniques (including CSP) for the downstream tasks of air temperature, median household income, elevation and population prediction shown in their work in \citep{satclip}, which highlights the model's ability to be utilized in diverse downstream tasks. We included GeoCLIP for its performance achieved in their work in limited-data settings using the MediaEval Placing Tasks 2016 dataset \citep{7849098,geoclip}; in the context of our work, we have a relatively small dataset of schools.

\section{Methodology}

\textbf{ML Classifiers:} We leverage shallow ML classifier architectures including Random Forest (RF),  Gradient Boosting (GB), Support Vector Machines (SVM), Logistic Regression (LR), Extreme Gradient Boosting (XGB) and the Multi-Layer Perception (MLP) neural network. We selected these models over more complex deep learning architecture due to the small size of our dataset and the history of superiority of tree-based models on tabular data over deep learning \citep{grinsztajn2022treebasedmodelsoutperformdeep}.

\begin{figure*}
    \centering
    {\setlength\intextsep{0pt}
    {\includegraphics[scale=0.2]{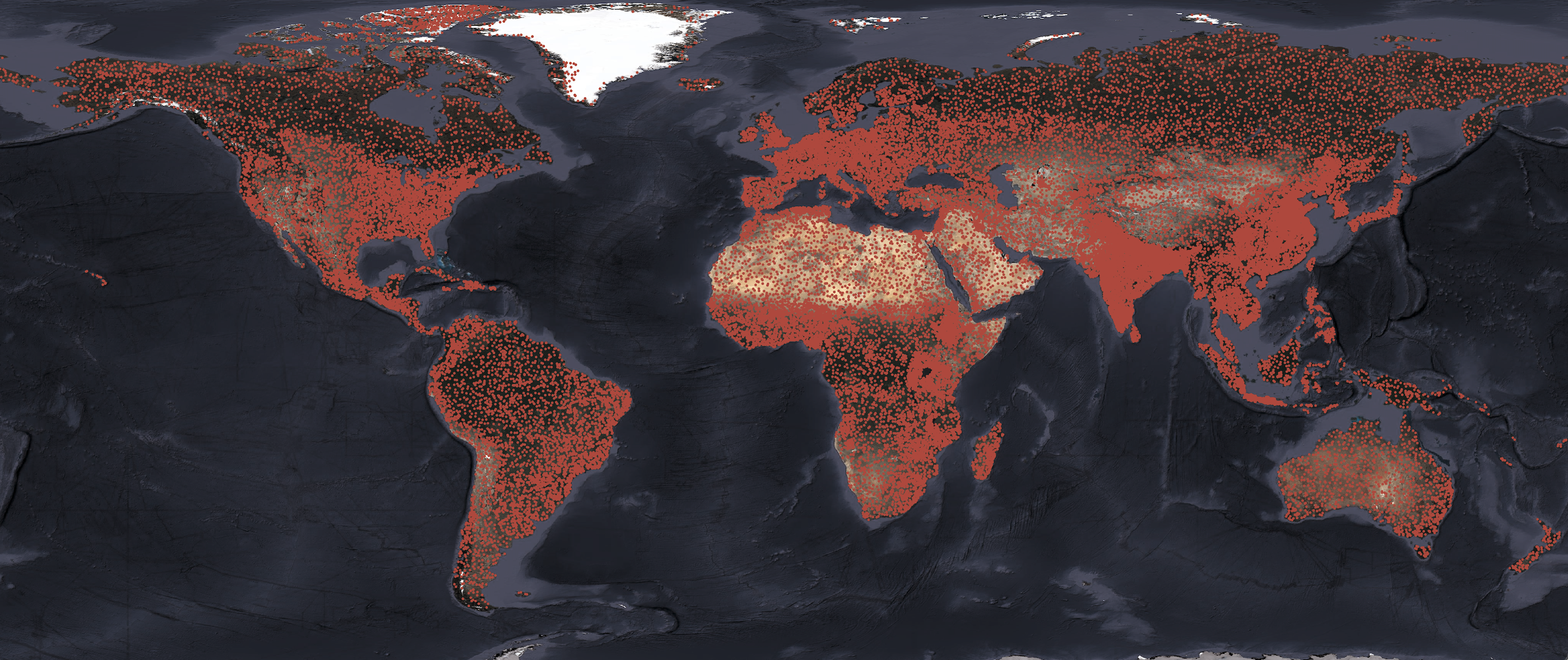}}
    \caption{PhilEO VHR sampling distribution
    \label{fig:esa}}}
\end{figure*}

\noindent\textbf{ESA PhilEO VHR:} The ESA $\phi$-lab PhilEO VHR Pre-cursor model is a deep learning architecture and semi-supervised pre-training methodology for training large-scale geographical aware deep learning models \citep{fibaek2024phileo}. The setup consists of a U-Net with learned skip-connection, where each block is penalized for transferring information of depending on spatial resolution of the connection. During training most of the skip connection are replaced with Gaussian noise of the same mean and standard deviation as the original as well as masking. Besides autoencoder features, the embedding space is used to attach a series of heads, which predict global features and coordinates of: Sinusoidal Encoded Coordinates \citep{fibaek2024phileo}, Ground Sampling Distance, Climate Zones \citep{Beck2018}, Land Cover \citep{zanaga2022}, Water Presence \citep{Melchiorri2023}, and Building Presence \citep{Melchiorri2023}.
A cosine-similarity based contrastive loss is applied on the embeddings versus embeddings of augmented images of the same location \citep{Wojke2018}. Owing to the global and large-scale dataset, the batch order can be flipped and the images used as negative sample pairs, although with less weight applied to negative samples. The sampling of the model is 50,000 global locations with settled areas being over sampled at twice the likelihood as other places, shown in Figure \ref{fig:esa}. Each city with at least 50,000 inhabitants is sampled at least once. Each location is sampled multiple times at different spatial resolution.

\noindent\textbf{Experimental Setup:}
We investigate four feature space setups for our prediction task: \begin{itemize}
    \item \textbf{Hand-crafted EO engineered features}: features derived from GEE, power grid, and speedtest data.
    \item \textbf{Feature embeddings}: vector outputs from location encoder models.
    \item \textbf{Combined}: combined engineered and PhilEO VHR Pre-cursor model embeddings.
    \item \textbf{Engineered features + auxiliary school data}: GEE, power grid, speedtest and ground-based survey data.
\end{itemize}

In all experiments, models are trained, validated and tested on a per-country level with a 70/15/15 train/validation/test split. We implement grid search cross validation to exhaustively tune model hyperparameters, with details of our defined search space included in our Supplementary Materials. Experimental setups one and two investigate model performance leveraging engineered features and extracted features from location encoders as inputs individually. Experiment three is designed to evaluate the hypothesis that the information learned from the location embeddings could be complementary to the engineered features, thereby increasing the overall performance of the models for connectivity prediction over using engineered or embeddings only. Experiment four investigates model performance when combining engineered features with ground-based survey data.

\section{Results and Discussion}
\begin{table*}
\centering
\small
\caption{Averaged test set F1 and Accuracy (Acc) using SatClip (S), GeoClip (Geo), CSP, features for Botswana (\textbf{BWA}) and Rwanda (\textbf{RWA}). }
\label{table:results}
\resizebox{\linewidth}{!}{%
\begin{tblr}{
  cell{1}{3} = {c=2}{},
  cell{1}{5} = {c=2}{},
  cell{1}{7} = {c=2}{},
  cell{1}{9} = {c=2}{},
  cell{1}{11} = {c=2}{},
  cell{1}{13} = {c=2}{},
  cell{1}{15} = {c=2}{c},
  cell{1}{17} = {c=2}{c},
  cell{3}{1} = {r=6}{},
  cell{9}{1} = {r=6}{},
  hline{2-3,9} = {2-18}{},
}
             &              & \textbf{S-R18-10} &             & \textbf{S-R18-40} &             & \textbf{S-R50-10} &             & \textbf{S-R50-40} &             & \textbf{S-V16-10} &             & \textbf{\textbf{S-V16-40}} &             & \textbf{Geo} &             & \textbf{CSP} &             \\
             &              & \textbf{Acc}      & \textbf{F1} & \textbf{Acc}      & \textbf{F1} & \textbf{Acc}      & \textbf{F1} & \textbf{Acc}      & \textbf{F1} & \textbf{Acc}      & \textbf{F1} & \textbf{Acc}               & \textbf{F1} & \textbf{Acc} & \textbf{F1} & \textbf{Acc} & \textbf{F1} \\
{\rotatebox{90}{\textbf{{BWA}}}}  & \textbf{RF}  & 0.56              & 0.49        & 0.56              & 0.53        & 0.58              & 0.54        & 0.58              & 0.55        & 0.56              & 0.47        & 0.57                       & 0.54        & 0.63         & 0.68        & 0.58         & 0.53        \\
             & \textbf{MLP} & 0.50              & 0.40        & 0.51              & 0.38        & 0.48              & 0.28        & 0.50              & 0.34        & 0.49              & 0.32        & 0.50                       & 0.39        & 0.55         & 0.55        & 0.52         & 0.54        \\
             & \textbf{GB}  & 0.57              & 0.50        & 0.52              & 0.50        & 0.52              & 0.43        & 0.54              & 0.44        & 0.54              & 0.45        & 0.57                       & 0.54        & 0.65         & 0.71        & 0.53         & 0.44        \\
             & \textbf{SVM} & 0.53              & 0.54        & 0.61              & 0.57        & 0.54              & 0.52        & 0.62              & 0.57        & 0.50              & 0.50        & 0.19                       & 0.50        & 0.54         & 0.70        & 0.52         & 0.48        \\
             & \textbf{LR}  & 0.52              & 0.68        & 0.54              & 0.45        & 0.51              & 0.40        & 0.55              & 0.58        & 0.52              & 0.15        & 0.47                       & 0.27        & 0.61         & 0.67        & 0.51         & 0.59        \\
             & \textbf{XGB} & 0.60              & 0.54        & 0.62              & 0.59        & 0.53              & 0.45        & 0.60              & 0.59        & 0.53              & 0.41        & 0.53                       & 0.40        & 0.63         & 0.69        & 0.55         & 0.52        \\
{\rotatebox{90}{\textbf{{RWA}}}} & \textbf{RF}  & 0.63              & 0.68        & 0.62              & 0.67        & 0.63              & 0.68        & 0.63              & 0.68        & 0.63              & 0.68        & 0.62                       & 0.68        & 0.62         & 0.69        & 0.60         & 0.65        \\
             & \textbf{MLP} & 0.60              & 0.66        & 0.60              & 0.67        & 0.62              & 0.62        & 0.61              & 0.63        & 0.60              & 0.61        & 0.57                       & 0.59        & 0.60         & 0.65        & 0.58         & 0.63        \\
             & \textbf{GB}  & 0.63              & 0.70        & 0.62              & 0.67        & 0.65              & 0.71        & 0.63              & 0.68        & 0.65              & 0.71        & 0.62                       & 0.68        & 0.63         & 0.69        & 0.59         & 0.66        \\
             & \textbf{SVM} & 0.64              & 0.71        & 0.54              & 0.70        & 0.54              & 0.70        & 0.54              & 0.70        & 0.59              & 0.69        & 0.54                       & 0.70        & 0.64         & 0.71        & 0.61         & 0.72        \\
             & \textbf{LR}  & 0.59              & 0.67        & 0.60              & 0.67        & 0.61              & 0.67        & 0.59              & 0.67        & 0.61              & 0.67        & 0.59                       & 0.67        & 0.64         & 0.69        & 0.60         & 0.70        \\
             & \textbf{XGB} & 0.65              & 0.71        & 0.65              & 0.72        & 0.63              & 0.69        & 0.63              & 0.71        & 0.63              & 0.69        & 0.64                       & 0.71        & 0.63         & 0.70        & 0.61         & 0.69        
\end{tblr}
}
\end{table*}

\begin{table}
\small
\centering
\caption{Averaged test set F1 and Accuracy (Acc) using Engineer (Eng), PhilEO (Phi) location encoder and combined features for Botswana (\textbf{BWA}) and Rwanda (\textbf{RWA}). }
\label{table:results2}
\begin{tblr}{
  cell{1}{3} = {c=2}{c},
  cell{1}{5} = {c=2}{c},
  cell{1}{7} = {c=2}{},
  cell{3}{1} = {r=6}{},
  cell{9}{1} = {r=6}{},
  hline{2-3,9} = {2-8}{},
}
             &              & \textbf{Eng}          &                       & \textbf{Phi} &             & \textbf{Phi+Eng}      &                       \\
             &              & \textbf{Acc}          & \textbf{F1}           & \textbf{Acc} & \textbf{F1} & \textbf{Acc}          & \textbf{F1}           \\
{\rotatebox{90}{\textbf{BWA}}} & \textbf{RF}  & 0.68                  & 0.71                  & 0.58         & 0.53        & 0.61                  & 0.71                  \\
             & \textbf{MLP} & 0.63                  & 0.66                  & 0.52         & 0.54        & 0.64                  & 0.69                  \\
             & \textbf{GB}  & 0.64                  & 0.67                  & 0.53         & 0.44        & \uline{\textbf{0.70}} & \uline{\textbf{0.73}} \\
             & \textbf{SVM} & 0.63                  & 0.68                  & 0.52         & 0.48        & 0.63                  & 0.68                  \\
             & \textbf{LR}  & 0.63                  & 0.67                  & 0.51         & 0.59        & 0.55                  & 0.58                  \\
             & \textbf{XGB} & 0.64                  & 0.67                  & 0.55         & 0.52        & 0.63                  & 0.68                  \\
{\rotatebox{90}{\textbf{RWA}}} & \textbf{RF}  & 0.63                  & 0.71                  & 0.55         & 0.69        & 0.56                  & 0.69                  \\
             & \textbf{MLP} & 0.59                  & 0.67                  & 0.52         & 0.62        & 0.53                  & 0.69                  \\
             & \textbf{GB}  & 0.61                  & 0.70                  & 0.53         & 0.63        & 0.56                  & 0.63                  \\
             & \textbf{SVM} & \textbf{\uline{0.62}} & \textbf{\uline{0.72}} & 0.55         & 0.70        & 0.56                  & 0.70                  \\
             & \textbf{LR}  & 0.59                  & 0.70                  & 0.55         & 0.68        & 0.55                  & 0.68                  \\
             & \textbf{XGB} & 0.61                  & 0.67                  & 0.54         & 0.66        & 0.59                  & 0.67                  
\end{tblr}
\end{table}

\textbf{Connectivity Prediction Model Performance:} For both Botswana and Rwanda, models trained with the engineered feature space out-perform the location-encoded feature spaces. In Botswana, the GB model trained with the PhilEO VHR Pre-cursor and engineered features slightly out-performs all other configurations, which suggests that our hypothesis that information learned from the location embeddings can be complementary to the engineered features, however we do not see these same results in Rwanda, with further discussions on the limitations of our methodology below. Tables \ref{table:results}-\ref{table:results2} summarize the F1 score and accuracy per classifier architecture and feature space. Overall, though feature embeddings have been claimed in the literature to sufficiently summarize complex socio-economic ground conditions from satellite imagery, summary statistics and metrics extracted using domain expert insight in our engineered feature space show greater performance, though not perfect, to our task of internet connectivity prediction. This is a particularly important takeaway when considering low-resource settings. It is unreasonable to assume a developing nation will have access to proprietary, high-resolution satellite imagery for infrastructure development; freely available data can act as a sufficient starting point for these socio-economic applications.

\begin{figure}[h]
\begin{tabular}{ll}
\includegraphics[width=0.47\linewidth]{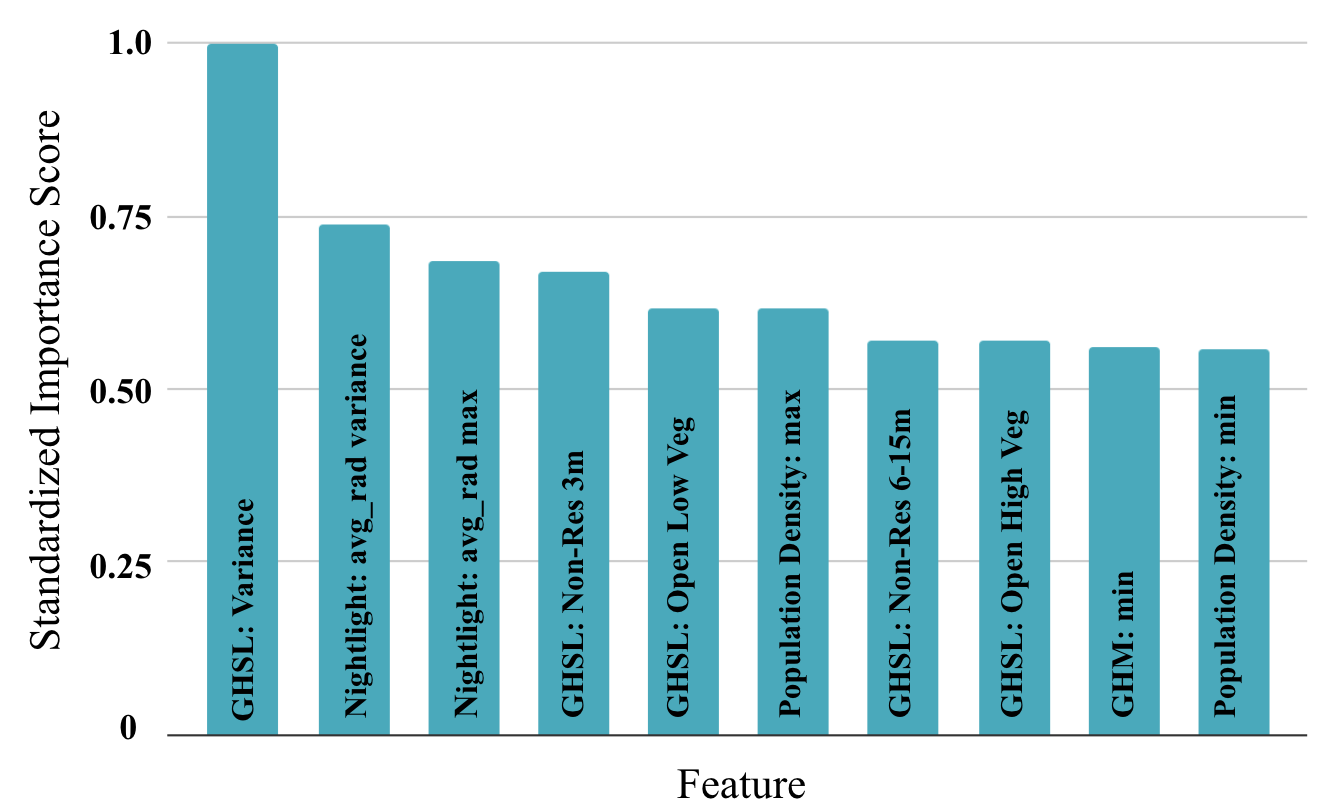}
&
\includegraphics[width=0.47\linewidth]{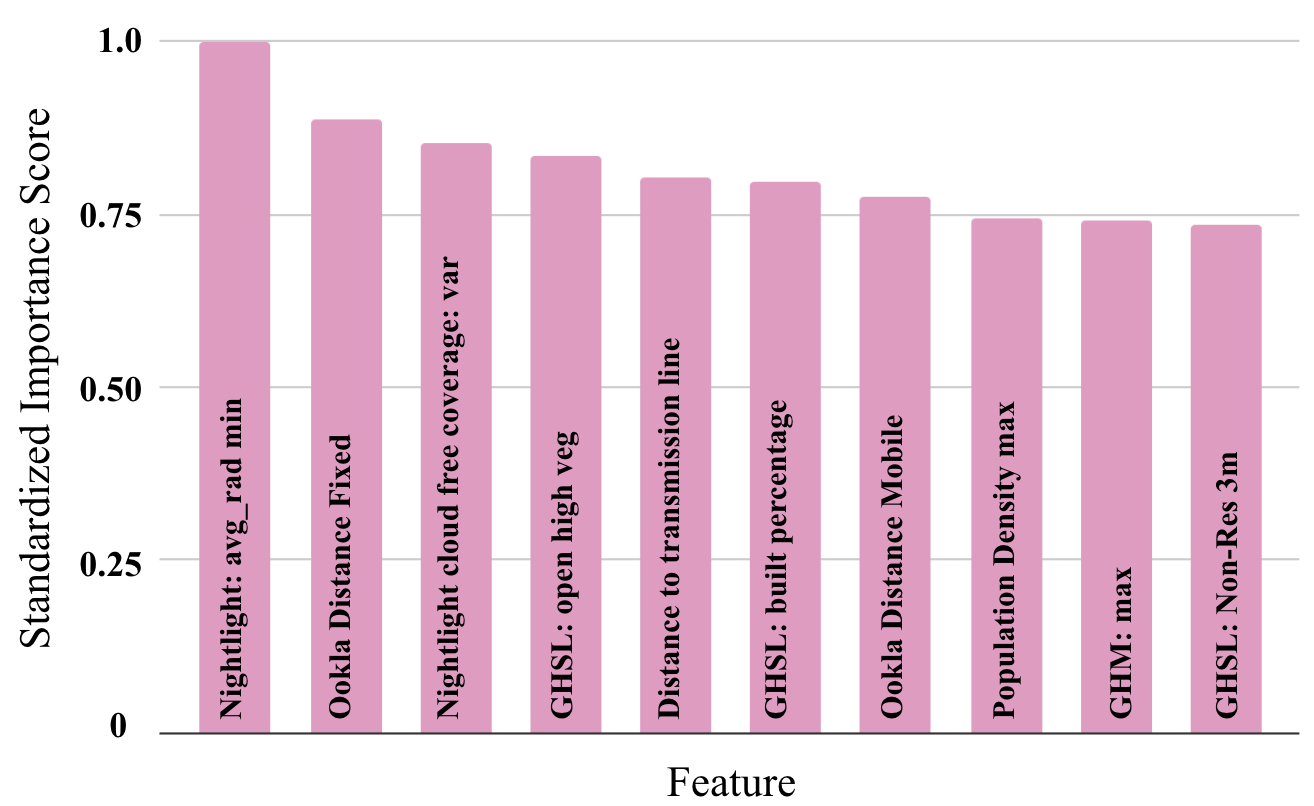}
\end{tabular}
\caption{Feature Importance. Left: Botswana Right: Rwanda}
\label{fig:FI}
\end{figure}

\noindent\textbf{Feature Importance:} Information from Global Human Settlement Layer (GHSL) strongly influenced the RF model's predictions and was present in four of the top ten features for Botswana and Rwanda. GHSL describes the inner characteristics of human settlements in terms of the functional and height-related components of the built environment. Intuitively, it is expected that the relative distance of a school to the power grid transmission line network would have a direct correlation to its electricity connection and therefore its internet connectivity. We see this claim supported in Rwanda in Figure \ref{fig:FI} where the distance to transmission line network is ranked 5th of the top ten features. Nighttime light features are also present in the top ten features for both countries. If we consider that internet connectivity prediction is related to poverty prediction (i.e., poor regions are less likely to be connected than higher-income regions), this result is consistent with the literature, whereby night-time light is used a proxy for poverty in numerous studies with ML for poverty prediction ~\citep{povproxy,XU2021102552,doi:10.1126/science.aaf7894}. \\

\noindent\textbf{Case Study: Kigali, Rwanda:}
\begin{figure*}
    \centering
    {\setlength\intextsep{0pt}
    {\includegraphics[scale=0.55]{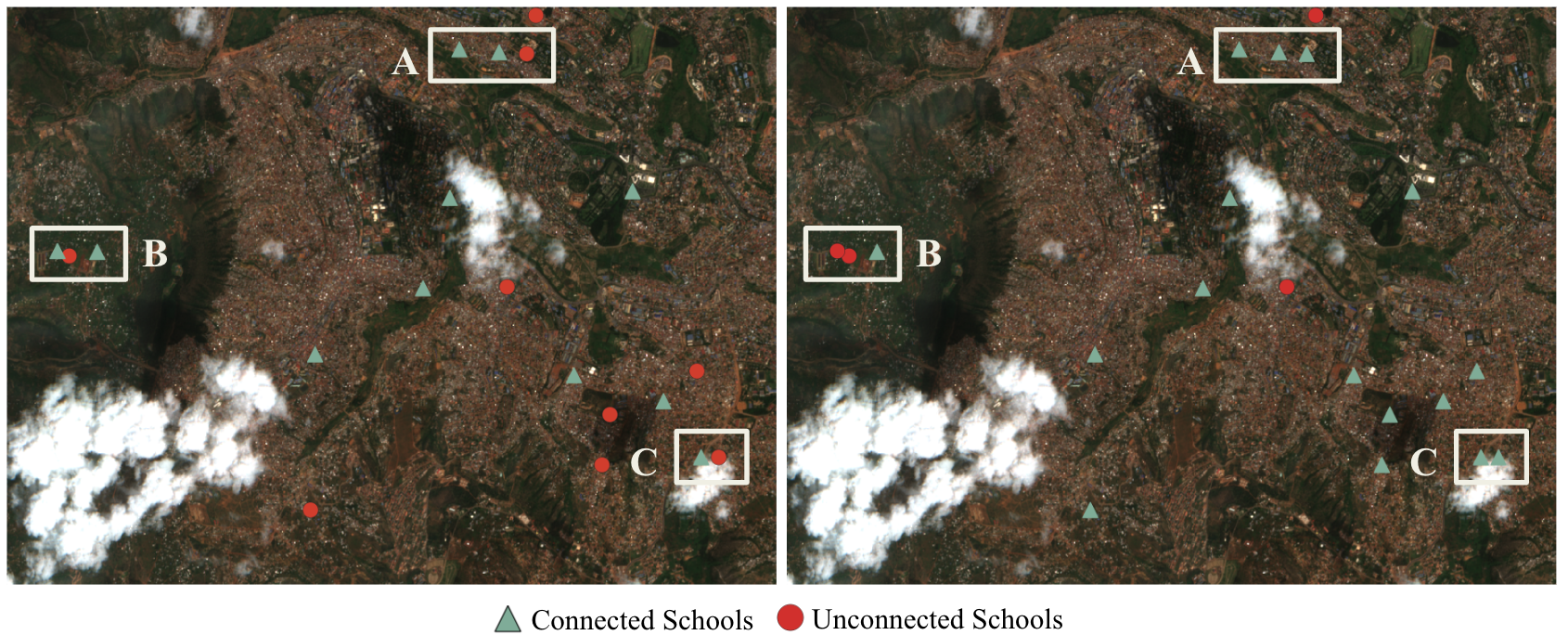}}
    \caption{South-West Kigali. \textbf{Left}: Ground-Truth labels, \textbf{Right}: Model Predictions.
    \label{fig:KIGALI}}}
\end{figure*}

To further investigate model performance, we take the case study of Kigali, Rwanda. We note that consistently, models using engineered features and location encoder-generated features struggle to perform greater than 62\% accuracy and 0.72 F1 score in the country, with the best performing models in terms of F1 score (SVM with engineered features) having an average False Positive (FP) rate of 63\%, True Positive (TP) rate of 87\%, False Negative (FN) rate of 13\%, and True Negative (TN) rate of 36\%. FPs are the most detrimental to real-world operationalization of our methodology, because this means that we are incorrectly identifying a school as connected, and would thereby be incorrectly diverting resources for improving digital infrastructure for these schools. To look further into why the FP rate is so high, we explore the ground truth conditions in the capital city of Kigali. We plot the connectivity ground truth and RF model predictions using the engineered feature space and highlight three locations in the South-Western part of the city (latitude, longitude bounds of (30.02,-2.01), (30.11,-1.93) respectively) of noteworthy ground truth distribution, shown in Figure \ref{fig:KIGALI}. We highlight three regions, A, B, and C, which show schools very close in proximity (less than 400m) from one another that have different connectivity status. In region A, the unconnected school is approximately 370m away from the nearest connected school, region B, the unconnected school is approximately 165m away from the nearest connected school, and for region C, the unconnected school is approximately 230m away from the nearest connected school. Unsurprisingly, our models struggle in these three regions; when considering our engineered feature space, this makes sense as we are taking data from a 1,000m radius with each school at the centre point, resulting in overlapping information between neighbouring schools that could be construing our model from delineating its true connectivity status.

\noindent\textbf{Incorporating Auxiliary School Information:} Across all model architectures and feature spaces, the models struggle with connectivity prediction accuracy, particularly over Rwanda. We therefore investigated incorporating additional auxiliary information about the schools in Experiment 4 using the UNICEF database, geoBoundaries\footnote{\tiny{\url{https://www.geoboundaries.org/countryDownloads.html}}} regional indicators, and Gridded Sex-Disaggregated School-Age Population \citep{schoolpop} into our feature space. From UNICEF, we added school education level, school distance to nearest Long-Term Evolution station, distance to nearest Universal Mobile Telecommunications System station, and distance to nearest Global System for Mobile communication station. We add regional indicators by treating the administrative boundaries of the country as categories and employ one-hot encoding on the categorical level-2 administrative zones of the country information to add as features. For edge-cases where the schools were slightly outside of a country's boundary (due to label noise), the administrative boundary that was closest to the school was taken as school's region. We incorporate the sum, mean, min, max, and variance of primary and secondary school age male and female students within a 1,000m radius of each school, with further details in Supplementary Materials. Results of our experiment are summarized in Table \ref{table:auxfeat}. The accuracy, F1 score and FP rate all noticeably improve when incorporating auxiliary school information. FP rates drop from 48\% to 19\% in Botswana and 64\% to 26\% in Rwanda, and accuracy scores reach 80\% and 73\% respectively.

\begin{table*}
\centering
\small
\caption{Averaged test set F1, FP and Accuracy (Acc) with auxiliary data for Botswana (\textbf{BWA}) and Rwanda (\textbf{RWA})}
\label{table:auxfeat}
{%
\begin{tblr}{
  row{1} = {c},
  cell{1}{1} = {c=7}{},
  cell{1}{8} = {c=6}{},
  cell{2}{2} = {c=3}{c},
  cell{2}{5} = {c=3}{c},
  cell{2}{8} = {c=3}{c},
  cell{2}{11} = {c=3}{c},
  vline{2} = {1}{},
  vline{3,6,9} = {2}{},
  vline{5,8,11} = {3-9}{},
  hline{1-2,4,10} = {-}{},
}
\textbf{BWA} &                       &                       &             &                          &                       &                       & \textbf{RWA}          &              &             &                          &                       &                       \\
             & \textbf{No Auxiliary} &                       &             & \textbf{Incl. Auxiliary} &                       &                       & \textbf{No Auxiliary} &              &             & \textbf{Incl. Auxiliary} &                       &                       \\
             & \textbf{F1}           & \textbf{\textbf{Acc}} & \textbf{FP} & \textbf{F1}              & \textbf{Acc}          & \textbf{FP}           & \textbf{F1}           & \textbf{Acc} & \textbf{FP} & \textbf{F1}              & \textbf{Acc}          & \textbf{FP}           \\
\textbf{RF}  & 0.71                  & 0.68                  & 0.40        & 0.77                     & 0.73                  & 0.39                  & 0.72                  & 0.63         & 0.65        & 0.73                     & 0.65                  & 0.62                  \\
\textbf{MLP} & 0.66                  & 0.63                  & 0.44        & 0.77                     & 0.74                  & 0.29                  & 0.67                  & 0.59         & 0.59        & 0.69                     & 0.66                  & 0.38                  \\
\textbf{GB}  & 0.67                  & 0.64                  & 0.39        & 0.79                     & 0.78                  & 0.19                  & 0.70                  & 0.61         & 0.67        & 0.71                     & 0.62                  & 0.62                  \\
\textbf{SVM} & 0.68                  & 0.63                  & 0.48        & 0.76                     & 0.75                  & 0.19                  & 0.72                  & 0.62         & 0.64        & \textbf{\uline{0.74}}    & \textbf{\uline{0.73}} & \textbf{\uline{0.26}} \\
\textbf{LR}  & 0.67                  & 0.63                  & 0.44        & 0.76                     & 0.69                  & 0.53                  & 0.70                  & 0.59         & 0.72        & 0.74                     & 0.71                  & 0.34                  \\
\textbf{XGB} & 0.67                  & 0.64                  & 0.40        & \textbf{\uline{0.81}}    & \textbf{\uline{0.80}} & \textbf{\uline{0.13}} & 0.67                  & 0.61         & 0.54        & 0.74                     & 0.68                  & 0.48                  
\end{tblr}}
\end{table*}

\section{Limitations and Future Directions}
\textbf{Globally Generalizable, Geographically-aware Models:} Our results show that classifier performance using the pre-trained location encoders as feature extractors is consistently limited across the four location encoder models. This is likely due to the complexity of our use-case, and relatively low spatial resolution of the data used to train the location encoders, and highlights the limitations of "globally-generalizable" location encoders for the downstream task of internet connectivity prediction.\\
\textbf{Label Quality:} A distinct limitation of this work is the overall quality of the dataset and the validity of the ground truth connectivity labels. As schools are connected by government partners, there is an undefined latency between connection and updated record of the connection shared, which could mean that there are schools whose connectivity status does not accurately represent its reality. To combat this, continued communication with country stakeholders is critical to maintain accurate, timely labels of school internet status.\\
\textbf{Connectivity Quality:} To maximize its impact on society, digital connectivity must be universal and meaningful, allowing users to have a safe, satisfying, enriching and productive online experience at an affordable cost \citep{connectivitymean}. Our methodology can be improved in future work to not only to know the binary (yes/no) connectivity status of a school, but to also estimate its connectivity \textit{quality}, through leveraging information like that available from Ookla Speedtest in addition to community partners who can monitor internet quality (i.e., download and upload speeds) in-schools. In addition, it would provide greater value to country stakeholders to incorporate a cluster-based analysis in our work, to include additional information about the socio-economic status of each connected/unconnected school. Future work will include collaborating with government partners to ground-validate model predictions.

\section{Conclusion}
Our work investigates Machine Learning for internet connectivity prediction in schools in Botswana and Rwanda, providing a new, multi-modal geospatial dataset and feature generation pipeline openly-available to the community, which can easily extend this study to other countries. We highlight the performance differences using hand-crafted features compared to geographically-aware location encoders, and show that incorporating auxiliary school information greatly improves predictive capabilities, achieving our highest accuracy, F1 scores of 80\%, 0.81 in Botswana and 73\%, 0.74 in Rwanda.

\bibliographystyle{plainnat}  
\bibliography{references.bib}  

\begin{thebibliography}{55}
\providecommand{\natexlab}[1]{#1}
\providecommand{\url}[1]{\texttt{#1}}
\expandafter\ifx\csname urlstyle\endcsname\relax
  \providecommand{\doi}[1]{doi: #1}\else
  \providecommand{\doi}{doi: \begingroup \urlstyle{rm}\Url}\fi

\bibitem[Arderne et~al.(2020)Arderne, Zorn, and Nicolas]{predelec}
C.~Arderne, C.~Zorn, and C.~et~al. Nicolas.
\newblock Predictive mapping of the global power system using open data.
\newblock \emph{Nature Scientific Data}, 7\penalty0 (19), 2020.
\newblock \doi{https://doi.org/10.1038/s41597-019-0347-4}.

\bibitem[atledge et~al.(2022)atledge, Cadamuro, and de~la Cuesta]{eleclive}
N.~atledge, G.~Cadamuro, and B.~et~al. de~la Cuesta.
\newblock Using machine learning to assess the livelihood impact of electricity access.
\newblock \emph{Nature}, 611:\penalty0 491--495, 2022.
\newblock \doi{10.1038/s41586-022-05322-8}.

\bibitem[Beck et~al.(2018)Beck, Zimmermann, McVicar, Vergopolan, Berg, and Wood]{Beck2018}
Hylke~E. Beck, Niklaus~E. Zimmermann, Tim~R. McVicar, Noemi Vergopolan, Alexis Berg, and Eric~F. Wood.
\newblock {Present and future Köppen-Geiger climate classification maps at 1-km resolution}.
\newblock 7 2018.
\newblock \doi{10.6084/m9.figshare.6396959}.
\newblock URL \url{https://figshare.com/articles/dataset/Present_and_future_K_ppen-Geiger_climate_classification_maps_at_1-km_resolution/6396959}.

\bibitem[Bon et~al.(2024)Bon, Saa-Dittoh, and Akkermans]{Bon2024}
Anna Bon, Francis Saa-Dittoh, and Hans Akkermans.
\newblock \emph{Bridging the Digital Divide}, pages 283--298.
\newblock Springer Nature Switzerland, Cham, 2024.
\newblock ISBN 978-3-031-45304-5.
\newblock \doi{10.1007/978-3-031-45304-5_19}.
\newblock URL \url{https://doi.org/10.1007/978-3-031-45304-5_19}.

\bibitem[Bondarenko et~al.(2022)Bondarenko, Sorichetta, Mesa, A, and Tatem]{schoolpop}
Maksym Bondarenko, Alessandro Sorichetta, Germ{\'a}n~Vargas Mesa, Am{\'e}lie~Gagnon A, and Andrew Tatem.
\newblock Gridded sex-disaggregated school-age population datasets for countries and dependent territories in africa in 2020, February 2022.
\newblock URL \url{https://eprints.soton.ac.uk/454468/}.

\bibitem[{Caribou Space}(2020)]{esa2030}
{Caribou Space}.
\newblock \emph{Adoption and Impact of Earth Observation for the 2030 Agenda for Sustainable Development}.
\newblock 2020.

\bibitem[Chen et~al.(2020)Chen, Kornblith, Norouzi, and Hinton]{simclr}
Ting Chen, Simon Kornblith, Mohammad Norouzi, and Geoffrey Hinton.
\newblock A simple framework for contrastive learning of visual representations.
\newblock In Hal~Daumé III and Aarti Singh, editors, \emph{Proceedings of the 37th International Conference on Machine Learning}, volume 119 of \emph{Proceedings of Machine Learning Research}, pages 1597--1607. PMLR, 13--18 Jul 2020.

\bibitem[{Christie} et~al.(2017){Christie}, {Fendley}, {Wilson}, and {Mukherjee}]{2017arXiv171107846C}
Gordon {Christie}, Neil {Fendley}, James {Wilson}, and Ryan {Mukherjee}.
\newblock {Functional Map of the World}.
\newblock \emph{arXiv e-prints}, art. arXiv:1711.07846, November 2017.
\newblock \doi{10.48550/arXiv.1711.07846}.

\bibitem[CIESIN(2018)]{popdata}
CIESIN.
\newblock Gridded population of the world, version 4 (gpwv4): Population density.
\newblock \emph{New York: NASA Socioeconomic Data and Applications Center (SEDAC)}, 2018.
\newblock \doi{https://doi.org/10.7927/H49C6VHW}.

\bibitem[Debien et~al.(2021)Debien, Casaburi, Milcinski, and Maranesi]{9553464}
Annekatrien Debien, Mauro Casaburi, Grega Milcinski, and Marcello Maranesi.
\newblock Esa's ai4eo initiative: Bridging the gap between the ai and earth observation communities.
\newblock In \emph{2021 IEEE International Geoscience and Remote Sensing Symposium IGARSS}, pages 251--253, 2021.
\newblock \doi{10.1109/IGARSS47720.2021.9553464}.

\bibitem[E et~al.(2021)E, J, T, I, V, Ishihara, B, and S]{generalmlsat}
Rolf E, Proctor J, Carleton T, Bolliger I, Shankar V, Ishihara, Recht B, and Hsiang S.
\newblock generalizable and accessible approach to machine learning with global satellite imagery.
\newblock \emph{Nature Communications}, 20\penalty0 (12), 2021.
\newblock \doi{10.1038/s41467-021-24638-z}.

\bibitem[Elvidge et~al.(2017)Elvidge, Baugh, Zhizhin, ChiHsu, and Ghosh]{nightlight}
Christopher~D. Elvidge, Kimgerly Baugh, Mikhail Zhizhin, Feng ChiHsu, and Tolottama Ghosh.
\newblock Viirs night-time lights.
\newblock \emph{International Journal of Remote Sensing}, 38\penalty0 (21):\penalty0 5860--5879, 2017.
\newblock \doi{https://doi.org/10.1080/01431161.2017.1342050}.

\bibitem[ESA(2020)]{esaEO}
ESA.
\newblock Newcomers earth observation guide, 2020.
\newblock URL \url{https://business.esa.int/newcomers-earth-observation-guide#:~:text=EO%20is%20defined%20as%20the,its%20own%20advantages%20and%20limitations.}

\bibitem[Fallatah et~al.(2020)Fallatah, Jones, and Mitchell]{fallatah2020object}
Ahmad Fallatah, Simon Jones, and David Mitchell.
\newblock Object-based random forest classification for informal settlements identification in the middle east: Jeddah a case study.
\newblock \emph{International Journal of Remote Sensing}, 41\penalty0 (11):\penalty0 4421--4445, 2020.

\bibitem[Fibaek et~al.(2024)Fibaek, Camilleri, Luyts, Dionelis, and Saux]{fibaek2024phileo}
Casper Fibaek, Luke Camilleri, Andreas Luyts, Nikolaos Dionelis, and Bertrand~Le Saux.
\newblock Phileo bench: Evaluating geo-spatial foundation models, 2024.

\bibitem[Fisher et~al.(2022)Fisher, Gibson, Liu, Abdar, Posa, Salimi-Khorshidi, Hassaine, Cai, Rahimi, and Mamouei]{rs14133072}
Thomas Fisher, Harry Gibson, Yunzhe Liu, Moloud Abdar, Marius Posa, Gholamreza Salimi-Khorshidi, Abdelaali Hassaine, Yutong Cai, Kazem Rahimi, and Mohammad Mamouei.
\newblock Uncertainty-aware interpretable deep learning for slum mapping and monitoring.
\newblock \emph{Remote Sensing}, 14\penalty0 (13), 2022.
\newblock ISSN 2072-4292.
\newblock \doi{10.3390/rs14133072}.
\newblock URL \url{https://www.mdpi.com/2072-4292/14/13/3072}.

\bibitem[Grinsztajn et~al.(2022)Grinsztajn, Oyallon, and Varoquaux]{grinsztajn2022treebasedmodelsoutperformdeep}
Léo Grinsztajn, Edouard Oyallon, and Gaël Varoquaux.
\newblock Why do tree-based models still outperform deep learning on tabular data?, 2022.
\newblock URL \url{https://arxiv.org/abs/2207.08815}.

\bibitem[Haack and Ryerson(2016)]{haack2016improving}
Barry Haack and Robert Ryerson.
\newblock Improving remote sensing research and education in developing countries: Approaches and recommendations.
\newblock \emph{International Journal of Applied Earth Observation and Geoinformation}, 45:\penalty0 77--83, 2016.

\bibitem[{Internet World Stats}(2023)]{worldinterent}
{Internet World Stats}.
\newblock Internet world stats - internet penetration in africa.
\newblock \url{https://www.internetworldstats.com/stats1.htm}, 2023.
\newblock [Accessed 20-02-2024].

\bibitem[{ITU}(2021)]{ITU}
{ITU}.
\newblock Measuring digital development: Facts and figures 2021.
\newblock \url{www.itu.int/en/ITU-D/Statistics/Pages/facts/default.aspx}, 2021.
\newblock [Accessed 20-02-2024].

\bibitem[{ITU and United Nations }(2022)]{connectivitymean}
{ITU and United Nations }.
\newblock Achieving universal and meaningful digital connectivity setting a baseline and targets for 2030.
\newblock \url{https://www.itu.int/itu-d/meetings/statistics/wp-content/uploads/sites/8/2022/04/UniversalMeaningfulDigitalConnectivityTargets2030_BackgroundPaper.pdf}, 2022.
\newblock [Accessed 20-02-2024].

\bibitem[Jean et~al.(2016)Jean, Burke, Xie, Davis, Lobell, and Ermon]{doi:10.1126/science.aaf7894}
Neal Jean, Marshall Burke, Michael Xie, W.~Matthew Davis, David~B. Lobell, and Stefano Ermon.
\newblock Combining satellite imagery and machine learning to predict poverty.
\newblock \emph{Science}, 353\penalty0 (6301):\penalty0 790--794, 2016.
\newblock \doi{10.1126/science.aaf7894}.
\newblock URL \url{https://www.science.org/doi/abs/10.1126/science.aaf7894}.

\bibitem[Jia and Benson(2020)]{10.1145/3394486.3403101}
Junteng Jia and Austion~R. Benson.
\newblock Residual correlation in graph neural network regression.
\newblock In \emph{Proceedings of the 26th ACM SIGKDD International Conference on Knowledge Discovery \& Data Mining}, KDD '20, page 588–598, New York, NY, USA, 2020. Association for Computing Machinery.
\newblock ISBN 9781450379984.
\newblock \doi{10.1145/3394486.3403101}.
\newblock URL \url{https://doi.org/10.1145/3394486.3403101}.

\bibitem[Kennedy et~al.(2019)Kennedy, Oakleaf, Theobald, Baurch-Murdo, and Kiesecker]{ghm}
C.M. Kennedy, J.R. Oakleaf, D.M. Theobald, S.~Baurch-Murdo, and J.~Kiesecker.
\newblock Managing the middle: A shift in conservation priorities based on the global human modification gradient.
\newblock \emph{Global Change Biology}, 00:1:\penalty0 1--16, 2019.
\newblock \doi{doi:10.1111/gcb.14549}.

\bibitem[{Klemmer} et~al.(2023){Klemmer}, {Rolf}, {Robinson}, {Mackey}, and {Ru{\ss}wurm}]{satclip}
Konstantin {Klemmer}, Esther {Rolf}, Caleb {Robinson}, Lester {Mackey}, and Marc {Ru{\ss}wurm}.
\newblock {SatCLIP: Global, General-Purpose Location Embeddings with Satellite Imagery}.
\newblock \emph{arXiv e-prints}, art. arXiv:2311.17179, November 2023.
\newblock \doi{10.48550/arXiv.2311.17179}.

\bibitem[Larson et~al.(2017)Larson, Soleymani, Gravier, Ionescu, and Jones]{7849098}
Martha Larson, Mohammad Soleymani, Guillaume Gravier, Bogdan Ionescu, and Gareth~J.F. Jones.
\newblock The benchmarking initiative for multimedia evaluation: Mediaeval 2016.
\newblock \emph{IEEE MultiMedia}, 24\penalty0 (1):\penalty0 93--96, 2017.
\newblock \doi{10.1109/MMUL.2017.9}.

\bibitem[M. and Panagiotis(2023)]{ghsl}
Pesaresi M. and P.~Panagiotis.
\newblock {GHS-BUILT-C R2023A - GHS Settlement Characteristics, derived from Sentinel2 composite (2018) and other GHS R2023A data}.
\newblock \emph{European Commission, Joint Research Centre (JRC)}, PID: http://data.europa.eu/89h/3c60ddf6-0586-4190-854b-f6aa0edc2a30, 2023.
\newblock \doi{doi:10.2905/3c60ddf6-0586-4190-854b-f6aa0edc2a30}.

\bibitem[{Mai} et~al.(2023){Mai}, {Lao}, {He}, {Song}, and {Ermon}]{csp}
Gengchen {Mai}, Ni~{Lao}, Yutong {He}, Jiaming {Song}, and Stefano {Ermon}.
\newblock {CSP: Self-Supervised Contrastive Spatial Pre-Training for Geospatial-Visual Representations}.
\newblock \emph{arXiv e-prints}, art. arXiv:2305.01118, May 2023.
\newblock \doi{10.48550/arXiv.2305.01118}.

\bibitem[Marcello et~al.(2023)Marcello, Michele, Martino, Panagiotis, Manuel, Luca, Pietro, Daniele, Katarzyna, Alessandra, Johannes, Pierpaolo, and Thomas]{JRC133256}
Schiavina Marcello, Melchiorri Michele, Pesaresi Martino, Politis Panagiotis, Carneiro Freire~Sergio Manuel, Maffenini Luca, Florio Pietro, Ehrlich Daniele, Goch Katarzyna, Carioli Alessandra, Uhl Johannes, Tommasi Pierpaolo, and Kemper Thomas.
\newblock {GHSL Data Package 2023}.
\newblock \penalty0 (KJ-03-23-103-EN-N (online)), 2023.
\newblock \doi{10.2760/098587 (online)}.

\bibitem[Melchiorri and Kemper(2023)]{Melchiorri2023}
Michele Melchiorri and Thomas Kemper.
\newblock Establishing an operational and continuous monitoring of global built-up surfaces with the copernicus global human settlement layer.
\newblock In \emph{2023 Joint Urban Remote Sensing Event (JURSE)}, pages 1--4, 2023.
\newblock \doi{10.1109/JURSE57346.2023.10144201}.

\bibitem[Noor et~al.(2008)Noor, Alegana, and Gething]{povproxy}
A.M. Noor, V.A. Alegana, and P.W. et~al Gething.
\newblock Using remotely sensed night-time light as a proxy for poverty in africa.
\newblock \emph{Population Health Metrics}, 6:\penalty0 5, 2008.
\newblock \doi{10.1186/1478-7954-6-5}.

\bibitem[{Ookla LLC.}(2023)]{ookla}
{Ookla LLC.}
\newblock Global fixed broadband and mobile network maps.
\newblock \url{https://www.ookla.com/ookla-for-good/open-data}, 2023.
\newblock [Accessed 16-01-2024].

\bibitem[Pace and Barry(1997)]{RePEc:eee:stapro:v:33:y:1997:i:3:p:291-297}
Kelley Pace and Ronald Barry.
\newblock Sparse spatial autoregressions.
\newblock \emph{Statistics and Probability Letters}, 33\penalty0 (3):\penalty0 291--297, 1997.
\newblock URL \url{https://EconPapers.repec.org/RePEc:eee:stapro:v:33:y:1997:i:3:p:291-297}.

\bibitem[Pettersson et~al.(2023)Pettersson, Kakooei, Ortheden, Johansson, and Daoud]{PetterssonKOJD23}
Markus~B. Pettersson, Mohammad Kakooei, Julia Ortheden, Fredrik~D. Johansson, and Adel Daoud.
\newblock Time series of satellite imagery improve deep learning estimates of neighborhood-level poverty in africa.
\newblock In \emph{Proceedings of the Thirty-Second International Joint Conference on Artificial Intelligence, {IJCAI} 2023, 19th-25th August 2023, Macao, SAR, China}, pages 6165--6173. ijcai.org, 2023.
\newblock \doi{10.24963/IJCAI.2023/684}.
\newblock URL \url{https://doi.org/10.24963/ijcai.2023/684}.

\bibitem[Putri et~al.(2023)Putri, Wijayanto, and Pramana]{PUTRI2023100889}
Salwa~Rizqina Putri, Arie~Wahyu Wijayanto, and Setia Pramana.
\newblock Multi-source satellite imagery and point of interest data for poverty mapping in east java, indonesia: Machine learning and deep learning approaches.
\newblock \emph{Remote Sensing Applications: Society and Environment}, 29:\penalty0 100889, 2023.
\newblock ISSN 2352-9385.
\newblock \doi{https://doi.org/10.1016/j.rsase.2022.100889}.
\newblock URL \url{https://www.sciencedirect.com/science/article/pii/S2352938522001975}.

\bibitem[Radford et~al.(2021{\natexlab{a}})Radford, Kim, Hallacy, Ramesh, Goh, Agarwal, Sastry, Askell, Mishkin, Clark, Krueger, and Sutskever]{clip}
Alec Radford, Jong~Wook Kim, Chris Hallacy, Aditya Ramesh, Gabriel Goh, Sandhini Agarwal, Girish Sastry, Amanda Askell, Pamela Mishkin, Jack Clark, Gretchen Krueger, and Ilya Sutskever.
\newblock Learning transferable visual models from natural language supervision.
\newblock \emph{CoRR}, abs/2103.00020, 2021{\natexlab{a}}.
\newblock URL \url{https://arxiv.org/abs/2103.00020}.

\bibitem[Radford et~al.(2021{\natexlab{b}})Radford, Kim, Hallacy, Ramesh, Goh, Agarwal, Sastry, Askell, Mishkin, Clark, Krueger, and Sutskever]{radford2021learning}
Alec Radford, Jong~Wook Kim, Chris Hallacy, Aditya Ramesh, Gabriel Goh, Sandhini Agarwal, Girish Sastry, Amanda Askell, Pamela Mishkin, Jack Clark, Gretchen Krueger, and Ilya Sutskever.
\newblock Learning transferable visual models from natural language supervision, 2021{\natexlab{b}}.

\bibitem[Shi et~al.(2020)Shi, Wang, and Fang]{shi2020artificial}
Zheyuan~Ryan Shi, Claire Wang, and Fei Fang.
\newblock Artificial intelligence for social good: A survey.
\newblock \emph{arXiv preprint arXiv:2001.01818}, 2020.

\bibitem[Sulla-Menashe and Friedl(2023)]{modislc}
Damien Sulla-Menashe and Mark~A Friedl.
\newblock Mcd12q1 modis/terra+aqua land cover type yearly l3 global 500m sin grid v006.
\newblock \emph{NASA EOSDIS Land Processes DAAC}, 2023.
\newblock \doi{https://doi.org/10.5067/MODIS/MCD12Q1.006}.

\bibitem[Tian et~al.(2022)Tian, Wu, Zeng, Watmough, Zhang, and Li]{TIAN20227}
Fuyou Tian, Bingfang Wu, Hongwei Zeng, Gary~R Watmough, Miao Zhang, and Yurui Li.
\newblock Detecting the linkage between arable land use and poverty using machine learning methods at global perspective.
\newblock \emph{Geography and Sustainability}, 3\penalty0 (1):\penalty0 7--20, 2022.
\newblock ISSN 2666-6839.
\newblock \doi{https://doi.org/10.1016/j.geosus.2022.01.001}.
\newblock URL \url{https://www.sciencedirect.com/science/article/pii/S2666683922000086}.

\bibitem[Tingzon et~al.(2019)Tingzon, Orden, Go, Sy, Sekara, Weber, Fatehkia, Garc\'{\i}a-Herranz, and Kim]{phil-pov}
I.~Tingzon, A.~Orden, K.~T. Go, S.~Sy, V.~Sekara, I.~Weber, M.~Fatehkia, M.~Garc\'{\i}a-Herranz, and D.~Kim.
\newblock Mapping poverty in the philippines using machine learning, satellite imagery, and crowd-sourced geospatial information.
\newblock \emph{The International Archives of the Photogrammetry, Remote Sensing and Spatial Information Sciences}, XLII-4/W19:\penalty0 425--431, 2019.
\newblock \doi{10.5194/isprs-archives-XLII-4-W19-425-2019}.
\newblock URL \url{https://isprs-archives.copernicus.org/articles/XLII-4-W19/425/2019/}.

\bibitem[{UNICEF}(2023{\natexlab{a}})]{Giga}
{UNICEF}.
\newblock Giga.
\newblock \url{{https://giga.global/}}, year = {2023}, note = {[Accessed 20-01-2024]},, 2023{\natexlab{a}}.

\bibitem[{UNICEF}(2023{\natexlab{b}})]{projconnect}
{UNICEF}.
\newblock Project connect.
\newblock \url{https://projectconnect.unicef.org/map/countries}, 2023{\natexlab{b}}.
\newblock [Accessed 30-11-2024].

\bibitem[{UNICEF Europe and Central Asia}(2020)]{unicefitu2024}
{UNICEF Europe and Central Asia}.
\newblock {T}wo thirds of the world’s school-age children have no internet access at home, new {U}{N}{I}{C}{E}{F}-{I}{T}{U} report says --- unicef.org.
\newblock \url{https://www.unicef.org/press-releases/two-thirds-worlds-school-age-children-have-no-internet-access-home-new-unicef-itu}, 2020.
\newblock [Accessed 31-01-2024].

\bibitem[Van~Dijk(2019)]{digitaldivide}
Jan~A.G.M. Van~Dijk.
\newblock \emph{Jan van Dijk (2019/2020) The Digital Divide. Polity Press}.
\newblock 10 2019.

\bibitem[{Van Horn} et~al.(2017){Van Horn}, {Mac Aodha}, {Song}, {Cui}, {Sun}, {Shepard}, {Adam}, {Perona}, and {Belongie}]{inaturalist}
Grant {Van Horn}, Oisin {Mac Aodha}, Yang {Song}, Yin {Cui}, Chen {Sun}, Alex {Shepard}, Hartwig {Adam}, Pietro {Perona}, and Serge {Belongie}.
\newblock {The iNaturalist Species Classification and Detection Dataset}.
\newblock \emph{arXiv e-prints}, art. arXiv:1707.06642, July 2017.
\newblock \doi{10.48550/arXiv.1707.06642}.

\bibitem[{Vivanco Cepeda} et~al.(2023){Vivanco Cepeda}, {Nayak}, and {Shah}]{geoclip}
Vicente {Vivanco Cepeda}, Gaurav~Kumar {Nayak}, and Mubarak {Shah}.
\newblock {GeoCLIP: Clip-Inspired Alignment between Locations and Images for Effective Worldwide Geo-localization}.
\newblock \emph{arXiv e-prints}, art. arXiv:2309.16020, September 2023.
\newblock \doi{10.48550/arXiv.2309.16020}.

\bibitem[Wang et~al.(2019)Wang, Kuffer, and Pfeffer]{wang2019role}
Jiong Wang, Monika Kuffer, and Karin Pfeffer.
\newblock The role of spatial heterogeneity in detecting urban slums.
\newblock \emph{Computers, environment and urban systems}, 73:\penalty0 95--107, 2019.

\bibitem[Wojke and Bewley(2018)]{Wojke2018}
Nicolai Wojke and Alex Bewley.
\newblock Deep cosine metric learning for person re-identification.
\newblock In \emph{2018 IEEE Winter Conference on Applications of Computer Vision (WACV)}, pages 748--756, 2018.
\newblock \doi{10.1109/WACV.2018.00087}.

\bibitem[Wurm et~al.(2019)Wurm, Stark, Zhu, Weigand, and Taubenb{\"o}ck]{wurm2019semantic}
Michael Wurm, Thomas Stark, Xiao~Xiang Zhu, Matthias Weigand, and Hannes Taubenb{\"o}ck.
\newblock Semantic segmentation of slums in satellite images using transfer learning on fully convolutional neural networks.
\newblock \emph{ISPRS journal of photogrammetry and remote sensing}, 150:\penalty0 59--69, 2019.

\bibitem[Xu et~al.(2021)Xu, Song, Li, Liu, and Cao]{XU2021102552}
Jianbin Xu, Jie Song, Baochao Li, Dan Liu, and Xiaoshu Cao.
\newblock Combining night time lights in prediction of poverty incidence at the county level.
\newblock \emph{Applied Geography}, 135:\penalty0 102552, 2021.
\newblock ISSN 0143-6228.
\newblock \doi{https://doi.org/10.1016/j.apgeog.2021.102552}.
\newblock URL \url{https://www.sciencedirect.com/science/article/pii/S0143622821001685}.

\bibitem[Yeh et~al.(2020)Yeh, Perez, and Driscoll]{pubsatafrica}
C.~Yeh, A.~Perez, and A.~et~al. Driscoll.
\newblock Using publicly available satellite imagery and deep learning to understand economic well-being in africa.
\newblock \emph{Nature Communications}, 11:\penalty0 2583, 2020.
\newblock \doi{10.1038/s41467-020-16185-w}.
\newblock URL \url{https://www.nature.com/articles/s41467-020-16185-w#citeas}.

\bibitem[Yin et~al.(2019)Yin, Liu, Zhang, Wang, Shah, and Zimmermann]{gps2vec}
Yifang Yin, Zhenguang Liu, Ying Zhang, Sheng Wang, Rajiv~Ratn Shah, and Roger Zimmermann.
\newblock Gps2vec: Towards generating worldwide gps embeddings.
\newblock In \emph{Proceedings of the 27th ACM SIGSPATIAL International Conference on Advances in Geographic Information Systems}, SIGSPATIAL '19, page 416–419, New York, NY, USA, 2019. Association for Computing Machinery.
\newblock ISBN 9781450369091.
\newblock \doi{10.1145/3347146.3359067}.
\newblock URL \url{https://doi.org/10.1145/3347146.3359067}.

\bibitem[Yu(2014)]{metadisc}
L.~Yu.
\newblock Meta-discoveries from a synthesis of satellite-based land-cover mapping research.
\newblock \emph{International Journal of Remote Sensing}, 35\penalty0 (4573), 2014.

\bibitem[Zanaga et~al.(2022)Zanaga, Van De~Kerchove, Daems, De~Keersmaecker, Brockmann, Kirches, Wevers, Cartus, Santoro, Fritz, Lesiv, Herold, Tsendbazar, Xu, Ramoino, and Arino]{zanaga2022}
Daniele Zanaga, Ruben Van De~Kerchove, Dirk Daems, Wanda De~Keersmaecker, Carsten Brockmann, Grit Kirches, Jan Wevers, Oliver Cartus, Maurizio Santoro, Steffen Fritz, Myroslava Lesiv, Martin Herold, Nandin-Erdene Tsendbazar, Panpan Xu, Fabrizio Ramoino, and Olivier Arino.
\newblock Esa worldcover 10 m 2021 v200, October 2022.
\newblock URL \url{https://doi.org/10.5281/zenodo.7254221}.

\end{thebibliography}

\newpage
\appendix
\section{Technical Appendix}

\section{Code and Data Availability}
We provide the entire project code with the documentation needed to generate the feature space, train, test and validate our models with links to the pre-processed and cleaned data to re-run experiments at: \textbf{https://github.com/kelsdoerksen/giga-connectivity}

\section{School Connectivity Dataset Creation}
\label{appendix:school-mapping-data}

\textbf{Geospatial Data Preprocessing.} To generate our geospatial features, we treat each school location (latitude, longitude point) with connectivity information as the center point for which we extract geospatial data surrounding the area of interest. To do so, we leverage the open-source datasets from Google Earth Engine (GEE), and specify a radius extent ranging from 300m, 500m, 750m, 1000m and 5000m, respectively. Using the airPy\footnote{\tiny{\url{https:/github.com/kelsdoerksen/airPy}}} data processing package, we modified the package to include support for the Global Human Settlement Layer (GHSL) and Global Human Modification (GHM) datasets, and extract GHSL, GHM, MODIS, VIIRS Nighttime Lights, and Gridded Population of the World data for each point in the varying array size respectively. We then calculate summary statistics from the image arrays including mode, mean, maximum, minimum, variance, and percent of land cover class where appropriate, to create our tabular dataset of geospatial features per school. Table \ref{tab:gee} summarizes the key information about the Google Earth Engine datasets used in our study.

\begin{table}[H]
\centering
\small
\caption{Google Earth Engine Datasets Used for Feature Space \label{tab:gee}}
\resizebox{\linewidth}{!}{
\begin{tblr}{
  cell{2}{2} = {r=2}{},
  cell{2}{3} = {r=2}{},
  cell{4}{2} = {r=5}{},
  cell{4}{3} = {r=5}{},
  cell{7}{1} = {r=2}{},
  cell{9}{1} = {r=7}{},
  cell{9}{2} = {r=7}{},
  cell{9}{3} = {r=7}{},
  cell{17}{2} = {r=2}{},
  cell{17}{3} = {r=2}{},
  cell{17}{4} = {r=2}{},
  hline{2,4,9,16-17,19} = {-}{},
}
\textbf{Dataset}                      & \textbf{Resolution} & \textbf{Period} & \textbf{Classes}                                                                                       \\
VIIRS Nighttime Day/Night~            & 463.83m             & 2022            & Average DNB radiance values                                                                            \\
Band Composites V1                    &                     &                 & Cloud-free coverages                                                                                   \\
                                      & 500m                & 2020            & Evergreen Needleleaf Forests,~Evergreen Broadleaf Forests,~Deciduous Needleleaf Forests,~              \\
MODIS Land Cover LC\_Type1            &                     &                 & Deciduous Broadleaf Forests,~Mixed Forests,~Closed Shrublands                                          \\
Yearly Global                         &                     &                 & Open Shrublands,~Woody Savannas,~Savannas, Grasslands                                                  \\
                                      &                     &                 & Permanent Wetlands,~Croplands,~Urban and Built-up Lands,~                                              \\
                                      &                     &                 & Cropland/Natural Vegetation Mosaics,~Permanent Snow and Ice,~Barren,~Water Bodies                      \\
Global Human Settlement Layer         & 10m                 & 2018            & open low vegetation surfaces,~open medium vegetation surfaces,~open high vegetation surfaces,          \\
                                      &                     &                 & open~ water surfaces, open road surfaces, built~residential - building height = 3m,                    \\
                                      &                     &                 & built residential - 3m  building height = 6m,~built residential - 6m  building height = 15m,           \\
                                      &                     &                 & built residential - 15m  building height = 30m, built residential - building height  30m,              \\
                                      &                     &                 & built~residential - building height = 3m,~built non-residential - 3m  building height = 6m,            \\
                                      &                     &                 & built non-residential - 6m  building height = 15m,~built non-residential - 15m  building height = 30m, \\
                                      &                     &                 & built non-residential - building height  30m                                                           \\
Global Human Modification~            & 1000m               & 2016            & global Human Modification index                                                                        \\
Population Density (Gridded~          & 927.67m             & 2020            & Estimated number of persons per square kilometer                                                       \\
Population of the World Version 4.11) &                     &                 &                                                                                                        
\end{tblr}
}
\end{table}

\begin{figure}[!htbp]
    \centering
    {\setlength\intextsep{0pt}
    {\includegraphics[scale=0.5]{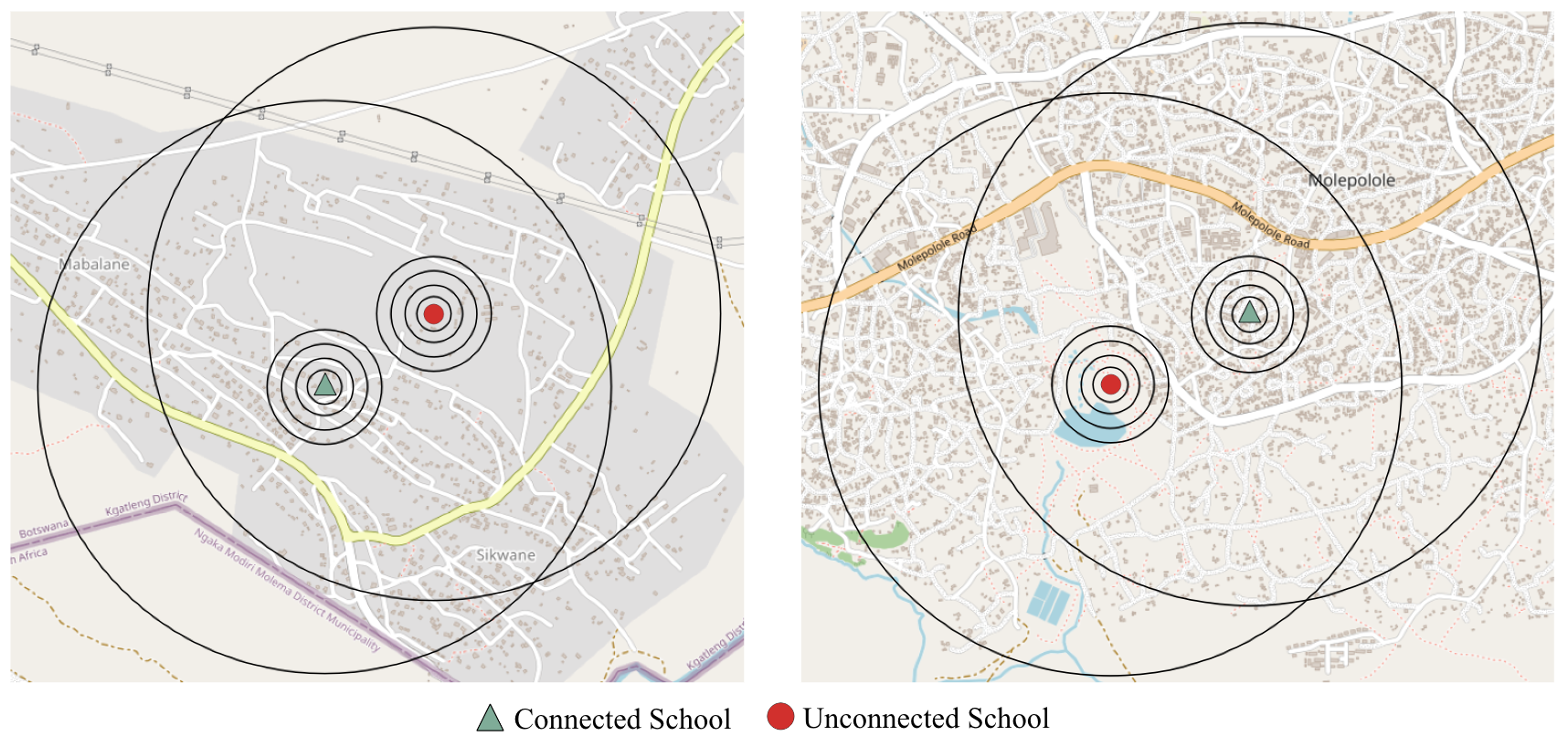}}
    \caption{Depiction of 300m, 500m, 750m, 1000m, and 5000m buffer extents surrounding connected (green) and unconnected (red) schools in Botswana in the Sikwane (left) and Molopolole (right) municipalities.  \label{fig:buffer_Exp}}}
\end{figure}

\begin{figure}[!htbp]
    \centering
    {\setlength\intextsep{0pt}
    {\includegraphics[scale=0.55]{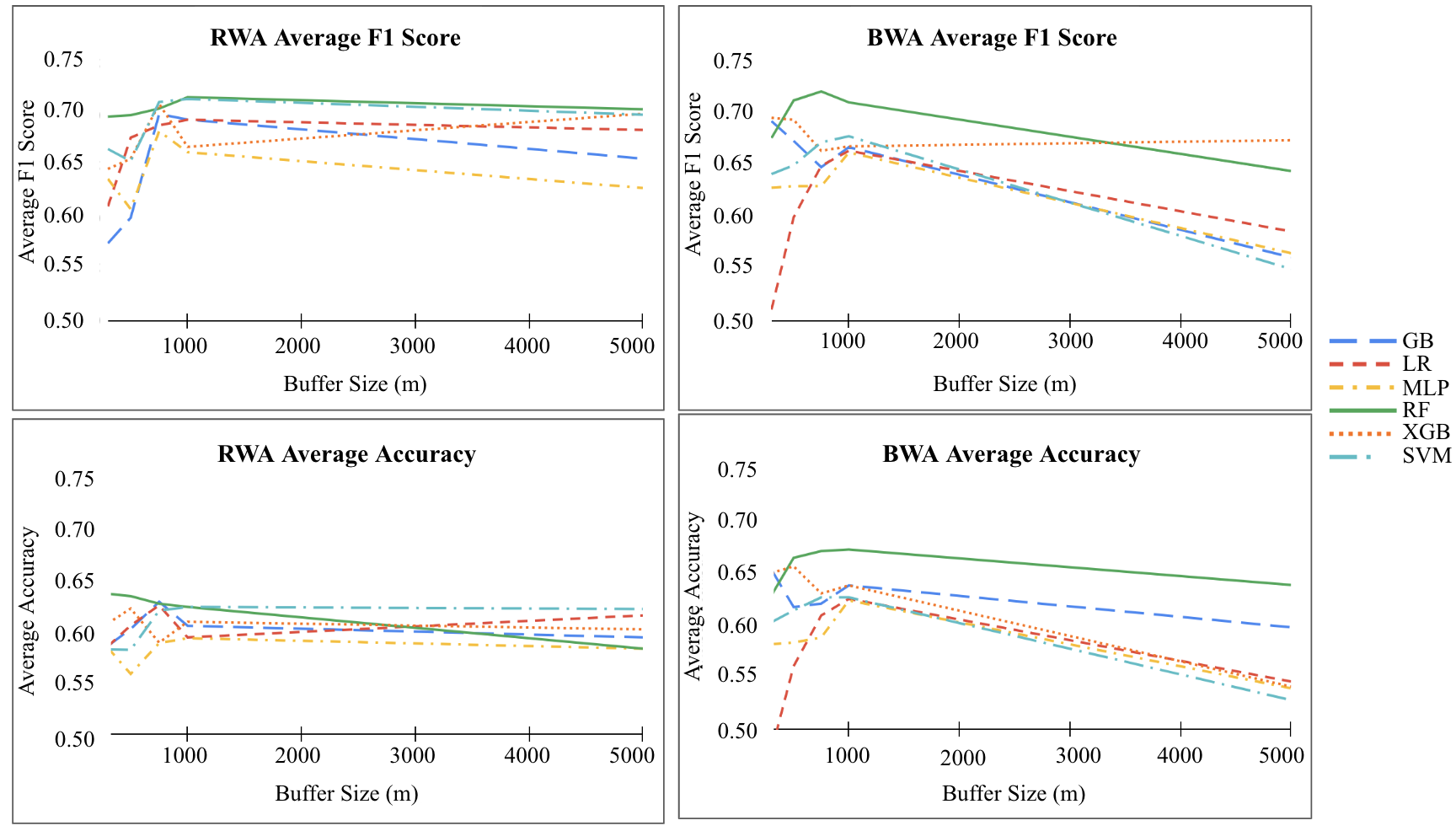}}
    \caption{Average Test Run Accuracy and F1-Score for varying ML classifier with varying buffer size in Rwanda (\textbf{RWA}) (left) and Botswana (\textbf{BWA}) (right)  \label{fig:buffer_plots}}}
\end{figure}

\textbf{Buffer extent selection.} We explore ML classifier performance across all model types for Botswana and Rwanda for varying buffer sizes of 300m, 500m, 750m, 1000m, and 5000m. The purpose of this experiment is to determine the optimal surrounding area to extract geospatial information for each school to create our feature space. Intuitively, too small of an area surrounding a school may not capture important information about the geographic context that could signal internet connectivity. For example, the Global Human Settlement Layer: Global settlement characteristics dataset used in our work includes information about global settlement characteristics at a 10m resolution. This includes classes such as open spaces and high vegetation surfaces which could indicate a school is in a rural area and perhaps less likely to be connected to electricity or the internet, versus a school that is surrounded by built spaces including residential and non-residential buildings of varying height, which could indicate more developed areas of the country that could be more likely to have internet infrastructure. Conversely, too large of an area surrounding the school could result in data leakage between neighbouring schools, which can result in model confusion and decrease in classification performance. We showcase examples of buffer overlap in Figure \ref{fig:buffer_Exp}, selecting two examples in the Molopolole and Sikwane municipalities in Botswana, we can see the connected (triangle) and unconnected (circle) schools have a overlap in their 5000m buffer extent, which could confuse our models. Table \ref{tab:geobuffer} and Figure \ref{fig:buffer_plots} summarize the average test set F1 and accuracy scores over varying buffer sizes for all classifiers. We can see that there is almost always a distinct decrease in model performance jumping from 1000m to 5000m buffer extent, with most models exhibiting the peak of their performance with the geospatial feature set between the 750m and 1000m buffer size. From these results, we select the 1000m buffer model's results to highlight in the paper and iterate with additional experimentation incorporating auxiliary school information.

\begin{table}
\centering
\small
\caption{Test Set Results for varying Feature Space with Geospatial and Movement data \label{tab:geobuffer}}
\label{tab:buffer_tab}{%
\begin{tblr}{
  cell{1}{3} = {c=3}{c},
  cell{1}{6} = {c=3}{c},
  cell{1}{9} = {c=3}{c},
  cell{1}{12} = {c=3}{c},
  cell{1}{15} = {c=2}{c},
  cell{3}{1} = {r=6}{},
  cell{3}{3} = {c},
  cell{3}{4} = {c},
  cell{3}{6} = {c},
  cell{3}{7} = {c},
  cell{3}{9} = {c},
  cell{3}{10} = {c},
  cell{3}{12} = {c},
  cell{3}{13} = {c},
  cell{3}{15} = {c},
  cell{3}{16} = {c},
  cell{4}{3} = {c},
  cell{4}{4} = {c},
  cell{4}{6} = {c},
  cell{4}{7} = {c},
  cell{4}{9} = {c},
  cell{4}{10} = {c},
  cell{4}{12} = {c},
  cell{4}{13} = {c},
  cell{4}{15} = {c},
  cell{4}{16} = {c},
  cell{5}{3} = {c},
  cell{5}{4} = {c},
  cell{5}{6} = {c},
  cell{5}{7} = {c},
  cell{5}{9} = {c},
  cell{5}{10} = {c},
  cell{5}{12} = {c},
  cell{5}{13} = {c},
  cell{5}{15} = {c},
  cell{5}{16} = {c},
  cell{6}{3} = {c},
  cell{6}{4} = {c},
  cell{6}{6} = {c},
  cell{6}{7} = {c},
  cell{6}{9} = {c},
  cell{6}{10} = {c},
  cell{6}{12} = {c},
  cell{6}{13} = {c},
  cell{6}{15} = {c},
  cell{6}{16} = {c},
  cell{9}{1} = {r=6}{},
  hline{2-3,9} = {2-16}{},
}
    &     & \textbf{300m} &      &  & \textbf{500m} &              &  & \textbf{750m} &              &  & \textbf{1000m} &              &  & \textbf{5000m} &      \\
    &     & Acc           & F1   &  & Acc           & F1           &  & Acc           & F1           &  & Acc            & F1           &  & Acc            & F1   \\
{\rotatebox{90}{\textbf{\footnotesize{{BWA}}}}} & RF  & 0.63          & 0.68 &  & \uline{0.67}  & \uline{0.72} &  & \uline{0.67}  & \uline{0.72} &  & \uline{0.68}   & \uline{0.71} &  & 0.64           & 0.65 \\
    & MLP & 0.58          & 0.63 &  & 0.59          & 0.63         &  & 0.59          & 0.64         &  & 0.63           & 0.66         &  & 0.54           & 0.57 \\
    & GB  & 0.66          & 0.7  &  & 0.62          & 0.68         &  & 0.62          & 0.65         &  & 0.64           & 0.67         &  & 0.60           & 0.56 \\
    & SVM & 0.60          & 0.64 &  & 0.62          & 0.65         &  & 0.65          & 0.67         &  & 0.63           & 0.68         &  & 0.53           & 0.55 \\
    & LR  & 0.48          & 0.51 &  & 0.56          & 0.60         &  & 0.61          & 0.65         &  & 0.63           & 0.67         &  & 0.55           & 0.59 \\
    & XGB & 0.65          & 0.70 &  & 0.66          & 0.70         &  & 0.63          & 0.67         &  & 0.64           & 0.67         &  & 0.54           & 0.68 \\
{\rotatebox{90}{\textbf{\footnotesize{{RWA}}}}} & RF  & 0.64          & 0.70 &  & 0.64          & 0.70         &  & 0.63          & 0.71         &  & 0.62           & 0.72         &  & 0.65           & 0.71 \\
    & MLP & 0.58          & 0.64 &  & 0.56          & 0.61         &  & 0.59          & 0.69         &  & 0.59           & 0.67         &  & 0.58           & 0.63 \\
    & GB  & 0.59          & 0.58 &  & 0.60          & 0.60         &  & 0.63          & 0.70         &  & 0.61           & 0.70         &  & 0.59           & 0.66 \\
    & SVM & 0.58          & 0.67 &  & 0.58          & 0.66         &  & 0.62          & 0.71         &  & 0.62           & 0.72         &  & 0.62           & 0.70 \\
    & LR  & 0.59          & 0.61 &  & 0.61          & 0.68         &  & 0.63          & 0.69         &  & 0.59           & 0.70         &  & 0.69           & 0.62 \\
    & XGB & 0.61          & 0.65 &  & 0.62          & 0.66         &  & 0.59          & 0.71         &  & 0.61           & 0.67         &  & 0.60           & 0.71 
\end{tblr}}
\end{table}

\textbf{Ookla Speedtest Data.} Ookla provides freely available global fixed broadband and mobile (cellular) network performance data of zoom level 16 web mercator tiles (approximately 610.8m x 610.8m at the equator). Download speed, upload speed and latency data are collected via Speedtest by Ookla applications for Androis and iOS and averaged for each tile, and measurements are filtered to results containing GPS-quality location accuracy. For each latitude, longitude point representing a school, we find the nearest Ookla tile and use the speedtest features defined in Table \ref{table:ookla} to represent the relevant fixed and broadband network performance of devices used around that school point.

\begin{table}[htbp]
\centering
\caption{Ookla SpeedTest Features \label{table:ookla}}
\resizebox{\linewidth}{!}{%
\begin{tabular}{l|l}
\textbf{Feature}        & \textbf{Description}                                                                                     \\ 
\hline
avg\_d\_kbps\_mobile    & Average download speed of all mobile tests performed in the nearest tile (kilobits per second)           \\
avg\_u\_kbps\_mobile    & Average upload speed of all mobile tests performed in the nearest tile (kilobits per second)             \\
avg\_lat\_ms\_mobile    & Average latency of all mobile tests performed in the nearest tile (milliseconds)                         \\
tests\_mobile           & Number of mobile tests taken in the nearest tile                                                         \\
devices\_mobile         & Number of unique mobile devices in the nearest tile                                                      \\
ookla\_distance\_mobile & Distance of the nearest tile to the school location                                                      \\
avg\_d\_kbps\_fixed     & Average download speed of all fixed broadband tests performed in the nearest tile (kilobits per second)  \\
avg\_u\_kbps\_fixed     & Average upload speed of all fixed broadband tests performed in the nearest tile (kilobits per second)    \\
avg\_lat\_ms\_fixed     & Average latency of all fixed broadband tests performed in the nearest tile (milliseconds)                \\
tests\_fixed            & Number of fixed broadband tests taken in the nearest tile                                                \\
devices\_fixed          & Number of unique fixed broadband devices in the nearest tile                                             \\
ookla\_distance\_fixed  & Distance of the nearest tile to the school location                                                     
\end{tabular}
}
\end{table}

\textbf{Electricity Transmission and Distribution Lines.} We use the derived map of global electricity transmission and distribution lines from the results of the gridfinder model, produced by ESMAP \footnote{\tiny{\url{https://zenodo.org/records/3266988/}}} \footnote{\tiny{\url{https://github.com/carderne/gridfinder/tree/1.0.0}}} to calculate the distance of each school to the nearest transmission line as a model feature. Gridfinder is an open-source tool that uses multiple filtering algorithms to night-time light imagery to identify locations most likely to be producing light from electricity, to create a globally consistent database of medium and lower voltage transmission lines \cite{predelec}. The tool has shown an accuracy of 75\% across 14 validation set countries, 9 of which are in Sub-Saharan Africa. To calculate the distance of each school to the nearest gridfinder transmission line, we use the geospatial software tool QGIS-LTR, and describe our steps for recreation in Table \ref{table:elec}.

\begin{table}
\centering
\caption{QGIS Data Processing Steps to Calculate School Distance to Nearest Transmission Line \label{table:elec}}
\resizebox{\linewidth}{!}{%
\begin{tblr}{
  vline{2-3} = {-}{},
  hline{2,4,7,9} = {-}{},
}
\textbf{Step} & \textbf{Instruction}                      & \textbf{QGIS-LTR Actions}                                                                                                                            \\
1             & Load World Bank Official Boundaries data~ & Select countries Layer $\rightarrow$ Open Attribute Table $\rightarrow$ Select Country of Interest $\rightarrow$                                                                         \\
              & and subset for country of interest        & Export Selected Layer as New Layer                                                                                                                   \\
              & Load Derived map of global electricity    & Select Transmission Line Layer $\rightarrow$ Vector Overlay: Extract/clip by extent $\rightarrow$                                                                            \\
2             & transmission and distribution lines gpkg  & Input layer: Transmission Line Layer, Extent: Country Layer                                                                                          \\
              & file and clip to AOI extent               &                                                                                                                                                      \\
3             & Load csv of school information as new~    & Layer $\rightarrow$ Add Layer $\rightarrow$ Add Delimited Text Layer                                                                                                           \\
              & School Locations Layer                    &                                                                                                                                                      \\
              &                                           & Select School Locations layer $\rightarrow$ Open Attribute Table $\rightarrow$ Toggle Editing $\rightarrow$                                                                              \\
4             & Calculate distance of school points to~   & New field (name: distance\_to\_transmission\_line) $\rightarrow$ Open field calculator $\rightarrow$                                                                         \\
              & nearest transmission line                 & Update existing field (distance\_to\_transmission\_line) $\rightarrow$~                                                                                          \\
              &                                           & length(make\_line (\textbackslash{}\$geometry,closest\_point (overlay\_nearest ('line', \textbackslash{}\$geometry)[0],\textbackslash{}\$geometry))) 
\end{tblr}
}
\end{table}

\textbf{Auxiliary Data: School-Age Population.} We explore the use of the Gridded Sex-Disaggregated School-Age Population (2020) dataset, which includes the sex disaggregated school age population for countries and Dependent territories in Africa. Because our work is focused particularly on internet connectivity in \textit{schools}, we add the male and female primary and secondary school age population sum, mean, mininmum, maximum, and variance surrounding each school location within a 1000m buffer. To generate the features, we take each latitude, longitude point representing a school and draw a 1000m radius buffer centered on the school. Using the zonal statistics functionality in QGIS-LTR, we calculate the summary statistics listed above in each buffer zone and use these as additional auxiliry information to train our models. The Gridded Sex-Disaggregated School-Age Population (2020) is freely available as .tiff files per country via the GEE Community Catalog \footnote{\tiny{\url{https://gee-community-catalog.org/projects/wpschool/}}}.

\textbf{Data Cleaning.}
We focus primarily on schools after early childhood education, removing schools that contain the keywords “preschool”, “nursery”, and “kindergarten”. Duplicate points (i.e. multiple points referring to the same school building), are removed by creating 25m buffers around each school location and grouping points that shared overlapping buffer zones into a single school. We identified schools with similar names by applying fuzzy string matching for pairs of schools that are within a distance of 300 m of each other. Using the RapidFuzz library\footnote{\tiny{\url{https://github.com/rapidfuzz/RapidFuzz}}}, we deduplicated school names by identifying strings that match with a normalized Levenshtein similarity of at least 85\% based on the optimal alignment of the two strings. Erroneous data points in uninhabited locations are removed by referencing open-source settlement data including the Microsoft building footprints\footnote{\tiny{\url{https://github.com/microsoft/GlobalMLBuildingFootprints}}} and GHSL \footnote{\tiny{\url{https://ghsl.jrc.ec.europa.eu/download.php?ds=builtC}}} \citep{JRC133256} datasets, drawing a 150m buffer around each point and calculating the sum of the pixels within the buffer area. Schools are retained only if locations where the pixel sum from the rasterized Microsoft building footprints and GHSL were both nonzero.

\section{Spatial Distribution of Samples}
\begin{figure*}[!htbp]
\centering
\begin{subfigure}{.52\textwidth}
  \centering
  \includegraphics[width=.74\linewidth]{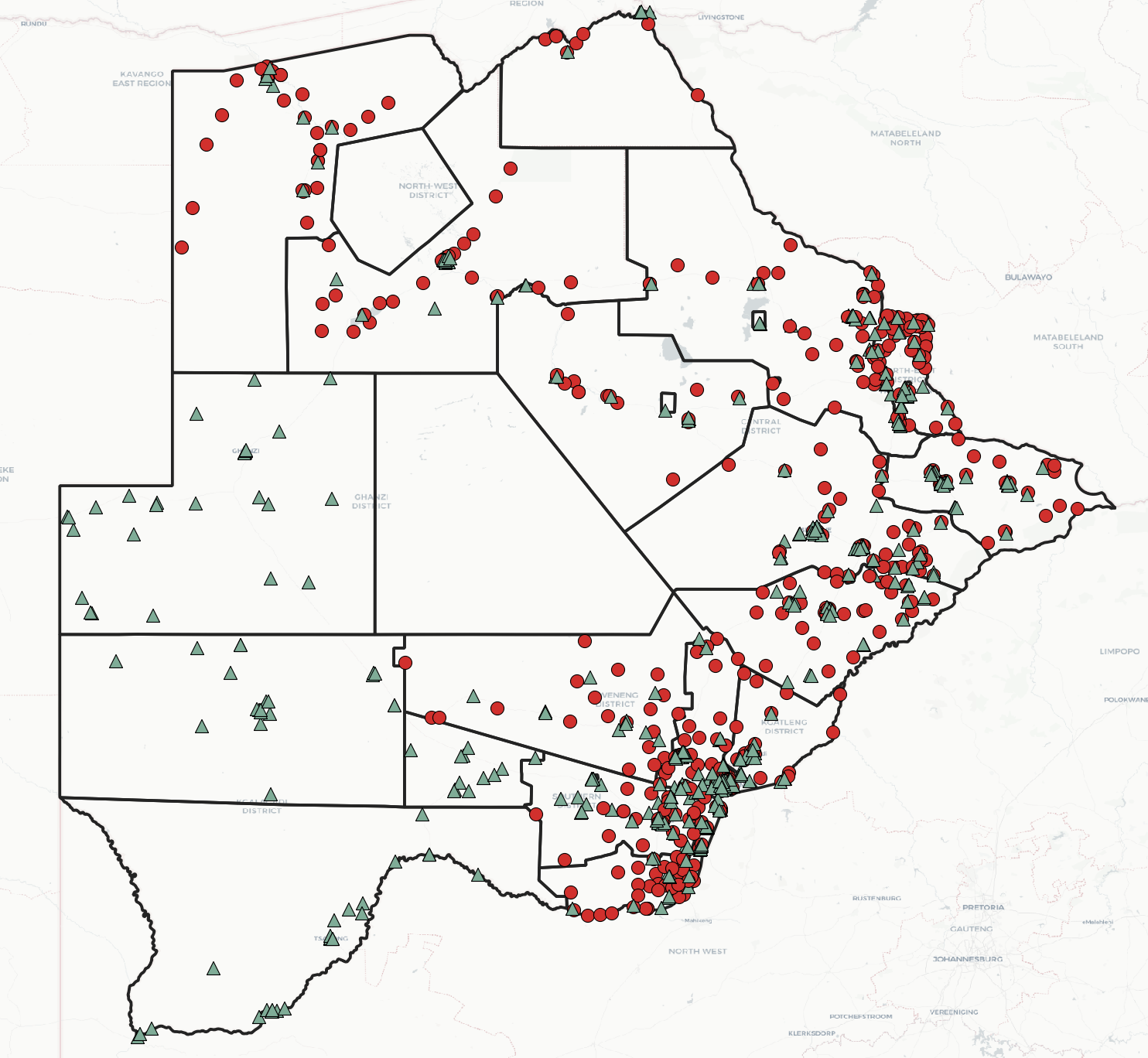}
  \caption{Botswana Connectivity Distribution}
  \label{fig:bwa_conn}
\end{subfigure}%
\begin{subfigure}{.52\textwidth}
  \centering
  \includegraphics[width=.8\linewidth]{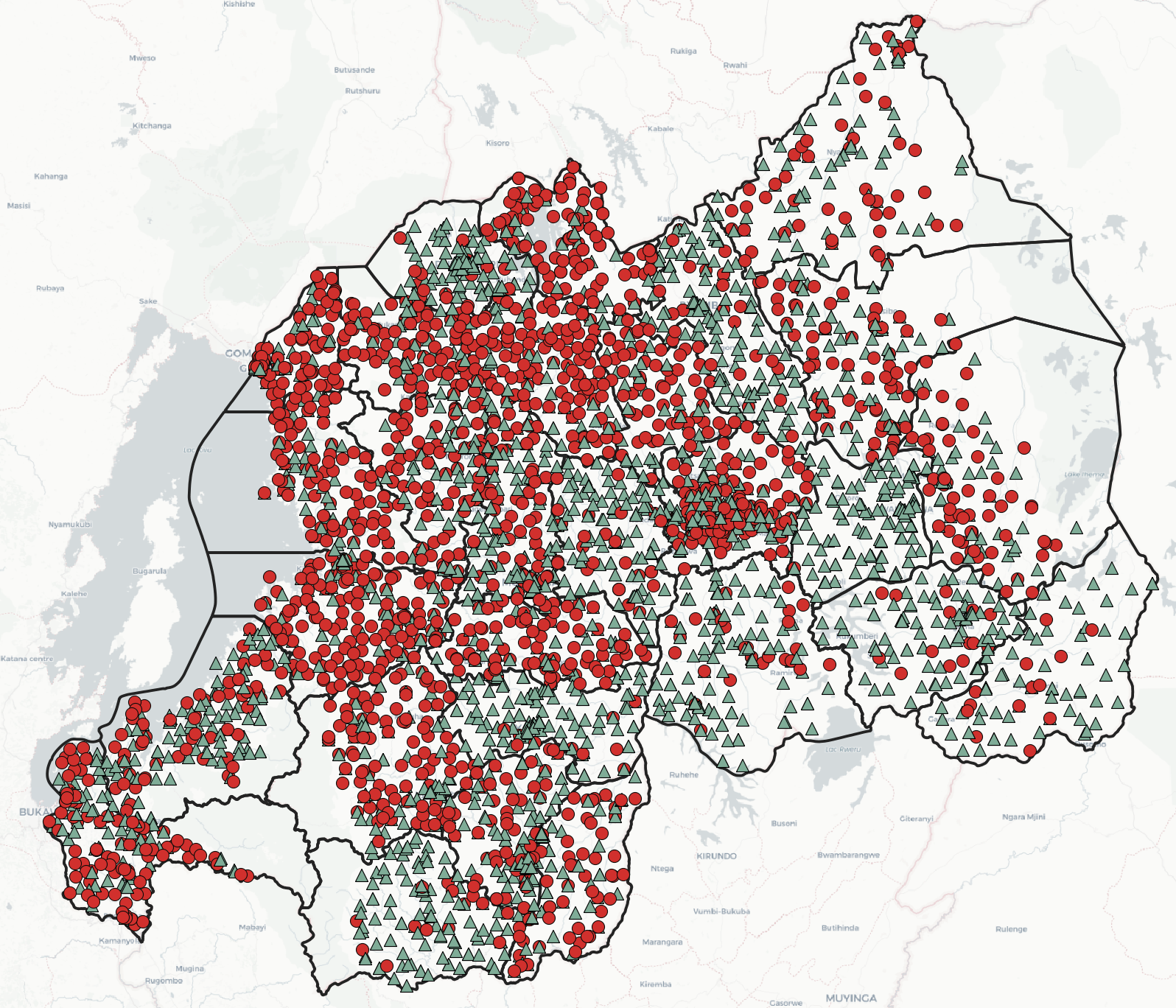}
  \caption{Rwanda Connectivity Distribution}
  \label{fig:sub2}
\end{subfigure}
\caption{Botswana (left) and Rwanda (right) spatial distribution of connected (green triangle) and unconnected (red circle) schools in each country.}
\label{spatial}
\end{figure*}

The spatial distribution of connected and unconnected schools in Botswana and Rwanda is shown in Figure \ref{spatial}. We can see that in Botswana, the density of schools is far lower than exhibited in Rwanda.

\section{Location Encoders}
\begin{table}[H]
\centering
\caption{Location encoder characteristics}
\small
\label{table:locencoders}{%
\begin{tblr}{
  hline{1-2} = {-}{},
}
\textbf{Name}       & \textbf{Dataset}            & \textbf{Embedding Size} \\
\textbf{SatCLIP}    & Sentinel-2                  & 256                     \\
\textbf{GeoCLIP}    & MediaEval Placing Tasks2016 & 512                     \\
\textbf{CSP}        & Functional Map of the World & 256                     \\
\textbf{PhilEO VHR} & VHR Collection              & 1024                    
\end{tblr}
}
\end{table}

To obtain the location embeddings for our school connectivity dataset from the pretrained models, we provide each school latitude, longitude coordinate to the location encoders during the model.eval)() stage and extract the corresponding embedding. Notebooks for the SatCLIP, GeoCLIP and CSP embedding generation are included in our shared git repository.

\textbf{SatCLIP.} The SatCLIP model uses the S2-100K dataset, which consists of 100,000 multi-spectral satellite images sampled from Sentinel-2 via the Microsoft Planetary Computer \footnote{\tiny{\url{https://planetarycomputer.microsoft.com/}}}. Images are sampled approximately uniformly over landmass, and include only images without cloud overage. We use all six pre-trained SatCLIP models in our study, trained with different vision encoders (ResNet, denoted by R, and Vision Transformer, denoted by V) and number of Legendre polynomials used for spherical harmonics location encoding (denoted by L): SatCLIP-ResNet18-L10, SatCLIP-ResNet18-L40, SatCLIP-ResNet50-L10, SatCLIP-ResNet50-L40, SatCLIP-ViT16-L10 and SatCLIP-ViT16-L40.

\textbf{GeoCLIP.}
GeoCLIP uses the MediaEval Placing Tasks 2016 (MP-16) dataset, which is composed of 4.7 million images taken across the globe. The GeoCLIP model consists of two main components; a Location encoder L(.) that obtains high-dimensional features from 2D GPS coordinates, and a CLIP-based image encoder V(.) to obtain the image features \cite{geoclip}. GeoCLIP uses two contrastive learning strategies, SimCLR and CLIP \citep{simclr,clip}. The pretrained model from CLIP is used as the backbone for the GeoCLIP image encoder, and the augmentation stratgey is similar to SimCLR, with the difference being that GeoCLIP matches an image to a gallery of GPS features.

\textbf{CSP.} The Contrastive Spatial Pre-Training (CSP) model uses the Functional Map of the World (fMoW) dataset, designed to inspire the devleopment of ML models capable of predicting the functional purpose of buildings and land use from temporal sequences of satellite images. The dataset contains over one million images from over 200 countries.

\section{Model Configurations and Setup}
Models are trained on a Intel(R) Core(TM) i9-10900K CPU /@ 3.70GHz Linux machine with 31 GB of available memory. Each model leverages hyperparameter tuning through a 5-fold grid search cross-validation scheme via scikit-learn on the validation set composed of 15\% of total available samples per country, with search spaces defined in Table \ref{table:params} and best model configuration, selected by maximizing the combined F1 score and accuracy, show in Table \ref{table:bestparams}.

\begin{table*}[h]
\small
\caption{School connectivity model parameter search spaces for Random Forest (\textbf{RF}) Support Vector Machines (\textbf{SVM}), Logistic Regression (\textbf{LR}), Gradient Boosting Classifier (\textbf{GB}), Extreme Gradient Boosting Classifier (\textbf{XGB}) and Multilayer Perceptron (\textbf{MLP}). \label{table:params}} 
\centering{
\begin{tabular}{ll}
\toprule
\textbf{Model} & \textbf{Parameters} \\
\midrule
\textbf{RF} & \makecell[l]{\textit{max depth}: 80, 90, 100  \\ \textit{max features}: 2, 3, 4  \\ \textit{min samples leaf}: 3, 4, 5  \\ \textit{min samples split}: 4, 6, 8  \\ \textit{n estimators}: 100, 200, 300, 500} \\
\midrule
\textbf{SVM} & \makecell[l]{\textit{C}: 0.001, 0.01, 0.1, 1.0, 10.0 \\ \textit{kernel}: linear, poly, rbf, sigmoid \\ \textit{degree}: 1, 2, 3, 4 \\ \textit{gamma}: scale, auto} \\
\midrule
\textbf{LR} & \makecell[l]{\textit{penalty}: l2, None \\ \textit{C}: 0.01, 0.1, 1.0}  \\
\midrule
\textbf{GB} & \makecell[l]{\textit{loss}: log loss, exponential \\ \textit{learning rate}: 0.05, 0.1, 0.5, 1 \\ \textit{n estimators}: 100, 200, 300 \\ \textit{criterion}: freidman-mse, squared-error \\ \textit{min samples split}: 2, 4, 6 \\ \textit{min samples leaf}: 1, 3, 5 \\ \textit{max features}: sqrt, log2, None}  \\
\midrule
\textbf{MLP} & \makecell[l]{\textit{hidden layer sizes}: (100,), (150,), (200,) \\ \textit{activation}: logistic, tanh, relu \\ \textit{solver}: lbfgs, sgd, adam \\ \textit{alpha}: 0.0001, 0.005, 0.001 \\ \textit{learning rate}: constant, invscaling, adaptive} \\
\midrule
\textbf{XGB} & \makecell[l]{\textit{eta}: 0.01, 0.05, 0.1, 0.15, 0.2 \\ \textit{max depth}: 3, 4, 5, 6 \\ \textit{min child weight}: 0, 1, 2 \\ \textit{max delta step}: 0, 1, 2 \\ \textit{subsample}: 0, 0.5, 0.75, 1  \\ \textit{tree method}: auto, exact, approx} \\
\bottomrule
\end{tabular}}
\end{table*}

\begin{table}
\centering
\caption{School connectivity model best parameter setting for Random Forest (\textbf{RF}) Support Vector Machines (\textbf{SVM}), Logistic Regression (\textbf{LR}), Gradient Boosting Classifier (\textbf{GB}), Extreme Gradient Boosting Classifier (\textbf{XGB}) and Multilayer Perceptron (\textbf{MLP}). \label{table:bestparams}}
{%
\begin{tblr}{
  cell{2}{1} = {r=5}{},
  cell{7}{1} = {r=4}{},
  cell{11}{1} = {r=2}{},
  cell{13}{1} = {r=7}{},
  cell{20}{1} = {r=5}{},
  cell{25}{1} = {r=6}{},
  hline{1,31} = {-}{0.08em},
  hline{2} = {1-2}{0.03em},
  hline{2} = {3-4}{},
  hline{7,11,13,20,25} = {-}{},
}
\textbf{Model} & \textbf{Parameters} & \textbf{BWA}  & \textbf{RWA}   \\
\textbf{RF}    & max depth           & 80            & 80             \\
               & max features        & 2             & 4              \\
               & min samples leaf    & 5             & 5              \\
               & min samples split   & 4             & 4              \\
               & n estimators        & 500           & 200            \\
\textbf{SVM}   & C                   & 0.1           & 0.1            \\
               & kernel              & linear        & poly           \\
               & degree              & 1             & 3              \\
               & gamma               & scale         & scale          \\
\textbf{LR}    & penalty             & l2            & l2             \\
               & C                   & 0.1           & 0.1            \\
\textbf{GB}    & loss                & exponential   & exponential    \\
               & learning rate       & 0.05          & 0.05           \\
               & n estimators        & 100           & 300            \\
               & criterion           & friedman\_mse & squared\_error \\
               & min samples split   & 2             & 2              \\
               & min samples leaf    & 5             & 3              \\
               & max features        & log2          & None           \\
\textbf{MLP}   & hidden layer sizes  & 200           & 200            \\
               & activation          & tanh          & relu           \\
               & solver              & sgd           & adam           \\
               & alpha               & 0.0001        & 0.005          \\
               & learning rate       & constant      & constant       \\
\textbf{XGB}   & eta                 & 0.01          & 0.2            \\
               & max depth           & 4             & 5              \\
               & min child weight    & 1             & 0              \\
               & max delta step      & 1             & 0              \\
               & subsample           & 0.75          & 1              \\
               & tree method         & exact         & exact          
\end{tblr}
}
\end{table}

\section{Additional Distributional Results}
We run each model architecture, per country, per feature space, five times and report the average performance in all tables in our main paper. We include the F1 and accuracy variance per experimental setup in Tables \ref{tab:rf_var} - \ref{tab:xgb_var} for Botswana and Tables \ref{tab:rf_rwa_var} - \ref{tab:xgb_rwa_var} for Rwanda below.

\begin{table}[H]
\centering
\begin{tblr}{
  cell{2}{3} = {c},
  cell{2}{4} = {c},
  cell{2}{5} = {c},
  cell{2}{6} = {c},
  cell{3}{3} = {c},
  cell{3}{4} = {c},
  cell{3}{5} = {c},
  cell{3}{6} = {c},
  cell{4}{3} = {c},
  cell{4}{4} = {c},
  cell{4}{5} = {c},
  cell{4}{6} = {c},
  cell{5}{3} = {c},
  cell{5}{4} = {c},
  cell{5}{5} = {c},
  cell{5}{6} = {c},
  cell{6}{3} = {c},
  cell{6}{4} = {c},
  cell{6}{5} = {c},
  cell{6}{6} = {c},
  cell{7}{3} = {c},
  cell{7}{4} = {c},
  cell{7}{5} = {c},
  cell{7}{6} = {c},
  cell{8}{3} = {c},
  cell{8}{4} = {c},
  cell{8}{5} = {c},
  cell{8}{6} = {c},
  cell{9}{3} = {c},
  cell{9}{4} = {c},
  cell{9}{5} = {c},
  cell{9}{6} = {c},
  cell{10}{3} = {c},
  cell{10}{4} = {c},
  cell{10}{5} = {c},
  cell{10}{6} = {c},
  cell{11}{3} = {c},
  cell{11}{4} = {c},
  cell{11}{5} = {c},
  cell{11}{6} = {c},
  cell{12}{3} = {c},
  cell{12}{4} = {c},
  cell{12}{5} = {c},
  cell{12}{6} = {c},
  vline{2} = {-}{},
  hline{1-2} = {-}{},
}
\textbf{Model} & \textbf{Features} & \textbf{Avg F1} & \textbf{F1 Variance} & \textbf{Avg Accuracy} & \textbf{Accuracy Variance} \\
               & S-R18-10          & 0.49            & 0.0006               & 0.56                  & 0.0002               \\
               & S-R18-40          & 0.53            & 0.0006               & 0.56                  & 0.0005               \\
               & S-R50-10          & 0.54            & 0.0004               & 0.58                  & 0.0001               \\
               & S-R50-40          & 0.55            & 0.0019               & 0.58                  & 0.0003               \\
               & S-V16-10          & 0.47            & 0.0025               & 0.56                  & 0.0007               \\
\textbf{RF}    & S-V16-40          & 0.54            & 0.0008               & 0.57                  & 0.0003               \\
               & GeoCLIP           & 0.63            & 0.0003               & 0.6                   & 0.0001               \\
               & CSP               & 0.53            & 0.0007               & 0.58                  & 0.0003               \\
               & Eng               & 0.71            & 0.0002               & 0.68                  & 0.0001               \\
               & Phi               & 0.53            &      0.0005                & 0.58                  &       0.0010               \\
               & Phi + Eng         & 0.71            & 0.0002               & 0.61                  & 0.0016               
\end{tblr}
\caption{Average F1 and Accuracy scores with variance over 5 runs for Random Forest Architecture for feature spaces SatClip (S-), GeoCLIP, CSP, Engineer (Eng), Phi and Combined (Phi + Eng), Botswana \label{tab:rf_var}}
\end{table}

\begin{table}[H]
\centering
\caption{Average F1 and Accuracy scores with variance over 5 runs for MLP Architecture for feature spaces SatClip (S-), GeoCLIP, CSP, Engineer (Eng), Phi and Combined (Phi + Eng), Botswana \label{tab:mlp_var}}
\small
\begin{tblr}{
  cell{2}{3} = {c},
  cell{2}{4} = {c},
  cell{2}{5} = {c},
  cell{2}{6} = {c},
  cell{3}{3} = {c},
  cell{3}{4} = {c},
  cell{3}{5} = {c},
  cell{3}{6} = {c},
  cell{4}{3} = {c},
  cell{4}{4} = {c},
  cell{4}{5} = {c},
  cell{4}{6} = {c},
  cell{5}{3} = {c},
  cell{5}{4} = {c},
  cell{5}{5} = {c},
  cell{5}{6} = {c},
  cell{6}{3} = {c},
  cell{6}{4} = {c},
  cell{6}{5} = {c},
  cell{6}{6} = {c},
  cell{7}{3} = {c},
  cell{7}{4} = {c},
  cell{7}{5} = {c},
  cell{7}{6} = {c},
  cell{8}{3} = {c},
  cell{8}{4} = {c},
  cell{8}{5} = {c},
  cell{8}{6} = {c},
  cell{9}{3} = {c},
  cell{9}{4} = {c},
  cell{9}{5} = {c},
  cell{9}{6} = {c},
  cell{10}{3} = {c},
  cell{10}{4} = {c},
  cell{10}{5} = {c},
  cell{10}{6} = {c},
  cell{11}{3} = {c},
  cell{11}{4} = {c},
  cell{11}{5} = {c},
  cell{11}{6} = {c},
  cell{12}{3} = {c},
  cell{12}{4} = {c},
  cell{12}{5} = {c},
  cell{12}{6} = {c},
  vline{2} = {-}{},
  hline{1-2} = {-}{},
}
\textbf{Model} & \textbf{Features} & \textbf{Avg F1} & \textbf{F1 Variance} & \textbf{Avg Accuracy} & \textbf{Accuracy Variance} \\
               & S-R18-10          & 0.40            & 0.0213               & 0.50                  & 0.0007               \\
               & S-R18-40          & 0.38            & 0.0069               & 0.51                  & 0.0003               \\
               & S-R50-10          & 0.28            & 0.0024               & 0.48                  & 0.0000               \\
               & S-R50-40          & 0.34            & 0.0077               & 0.50                  & 0.0004               \\
               & S-V16-10          & 0.32            & 0.0136               & 0.49                  & 0.0006               \\
\textbf{MLP}   & S-V16-40          & 0.39            & 0.0177               & 0.50                  & 0.0013               \\
               & GeoCLIP           & 0.55            & 0.0095               & 0.55                  & 0.0040               \\
               & CSP               & 0.54            & 0.0023               & 0.52                  & 0.0012               \\
               & Eng               & 0.66            & 0.0001               & 0.63                  & 0.0002               \\
               & Phi               & 0.54            &        0.0025              & 0.52                  &     0.0005                 \\
               & Phi + Eng         & 0.69            & 0.0028               & 0.64                  & 0.0007               
\end{tblr}
\end{table}

\begin{table}[H]
\centering
\caption{Average F1 and Accuracy scores with variance over 5 runs for GB Architecture for feature spaces SatClip (S-), GeoCLIP, CSP, Engineer (Eng), Phi and Combined (Phi + Eng), Botswana \label{tab:gb_var}}
\begin{tblr}{
  cell{2}{3} = {c},
  cell{2}{4} = {c},
  cell{2}{5} = {c},
  cell{2}{6} = {c},
  cell{3}{3} = {c},
  cell{3}{4} = {c},
  cell{3}{5} = {c},
  cell{3}{6} = {c},
  cell{4}{3} = {c},
  cell{4}{4} = {c},
  cell{4}{5} = {c},
  cell{4}{6} = {c},
  cell{5}{3} = {c},
  cell{5}{4} = {c},
  cell{5}{5} = {c},
  cell{5}{6} = {c},
  cell{6}{3} = {c},
  cell{6}{4} = {c},
  cell{6}{5} = {c},
  cell{6}{6} = {c},
  cell{7}{3} = {c},
  cell{7}{4} = {c},
  cell{7}{5} = {c},
  cell{7}{6} = {c},
  cell{8}{3} = {c},
  cell{8}{4} = {c},
  cell{8}{5} = {c},
  cell{8}{6} = {c},
  cell{9}{3} = {c},
  cell{9}{4} = {c},
  cell{9}{5} = {c},
  cell{9}{6} = {c},
  cell{10}{3} = {c},
  cell{10}{4} = {c},
  cell{10}{5} = {c},
  cell{10}{6} = {c},
  cell{11}{3} = {c},
  cell{11}{4} = {c},
  cell{11}{5} = {c},
  cell{11}{6} = {c},
  cell{12}{3} = {c},
  cell{12}{4} = {c},
  cell{12}{5} = {c},
  cell{12}{6} = {c},
  vline{2} = {-}{},
  hline{1-2} = {-}{},
}
\textbf{Model} & \textbf{Features} & \textbf{Avg F1} & \textbf{F1 Variance} & \textbf{Avg Accuracy} & \textbf{Accuracy Variance} \\
               & S-R18-10          & 0.5             & 0.0012               & 0.57                  & 0.0007                     \\
               & S-R18-40          & 0.5             & 0.0034               & 0.52                  & 0.0013                     \\
               & S-R50-10          & 0.43            & 0.0064               & 0.52                  & 0.0018                     \\
               & S-R50-40          & 0.44            & 0.0005               & 0.54                  & 0.0003                     \\
               & S-V16-10          & 0.45            & 0.0005               & 0.54                  & 0.0014                     \\
\textbf{GB}    & S-V16-40          & 0.54            & 0.0037               & 0.57                  & 0.0000                     \\
               & GeoCLIP           & 0.59            & 0.0039               & 0.56                  & 0.0047                     \\
               & CSP               & 0.44            & 0.0017               & 0.53                  & 0.0007                     \\
               & Eng               & 0.67            & 0.0004               & 0.64                  & 0.0004                     \\
               & Phi               & 0.44            &        0.0020              & 0.53                  &       0.0003                     \\
               & Phi + Eng         & 0.73            & 0.0004               & 0.7                   & 0.0014                     
\end{tblr}
\end{table}

\begin{table}[H]
\centering
\caption{Average F1 and Accuracy scores with variance over 5 runs for SVM Architecture for feature spaces SatClip (S-), GeoCLIP, CSP, Engineer (Eng), Phi and Combined (Phi + Eng), Botswana \label{svm_var}}
\begin{tblr}{
  cell{2}{3} = {c},
  cell{2}{4} = {c},
  cell{2}{5} = {c},
  cell{2}{6} = {c},
  cell{3}{3} = {c},
  cell{3}{4} = {c},
  cell{3}{5} = {c},
  cell{3}{6} = {c},
  cell{4}{3} = {c},
  cell{4}{4} = {c},
  cell{4}{5} = {c},
  cell{4}{6} = {c},
  cell{5}{3} = {c},
  cell{5}{4} = {c},
  cell{5}{5} = {c},
  cell{5}{6} = {c},
  cell{6}{3} = {c},
  cell{6}{4} = {c},
  cell{6}{5} = {c},
  cell{6}{6} = {c},
  cell{7}{3} = {c},
  cell{7}{4} = {c},
  cell{7}{5} = {c},
  cell{7}{6} = {c},
  cell{8}{3} = {c},
  cell{8}{4} = {c},
  cell{8}{5} = {c},
  cell{8}{6} = {c},
  cell{9}{3} = {c},
  cell{9}{4} = {c},
  cell{9}{5} = {c},
  cell{9}{6} = {c},
  cell{10}{3} = {c},
  cell{10}{4} = {c},
  cell{10}{5} = {c},
  cell{10}{6} = {c},
  cell{11}{3} = {c},
  cell{11}{4} = {c},
  cell{11}{5} = {c},
  cell{11}{6} = {c},
  cell{12}{3} = {c},
  cell{12}{4} = {c},
  cell{12}{5} = {c},
  cell{12}{6} = {c},
  vline{2} = {-}{},
  hline{1-2} = {-}{},
}
\textbf{Model} & \textbf{Features} & \textbf{Avg F1} & \textbf{F1 Variance} & \textbf{Avg Accuracy} & \textbf{Accuracy Variance} \\
               & S-R18-10          & 0.54            & 0.0000               & 0.53                  & 0.0000                     \\
               & S-R18-40          & 0.57            & 0.0000               & 0.61                  & 0.0000                     \\
               & S-R50-10          & 0.52            & 0.0003               & 0.54                  & 0.0001                     \\
               & S-R50-40          & 0.57            & 0.0000               & 0.62                  & 0.0000                     \\
               & S-V16-10          & 0.5             & 0.0000               & 0.5                   & 0.0000                     \\
\textbf{SVM}   & S-V16-40          & 0.5             & 0.0000               & 0.19                  & 0.0000                     \\
               & GeoCLIP           & 0.58            & 0.0002               & 0.57                  & 0.0001                     \\
               & CSP               & 0.48            & 0.0000               & 0.52                  & 0.0000                     \\
               & Eng               & 0.68            & 0.0002               & 0.63                  & 0.0001                     \\
               & Phi               & 0.48            &     0.0000                 & 0.52                  &      0.0001                      \\
               & Phi + Eng         & 0.68            & 0.0001               & 0.63                  & 0.0001                     
\end{tblr}
\end{table}

\begin{table}[H]
\centering
\caption{Average F1 and Accuracy scores with variance over 5 runs for LR Architecture for feature spaces SatClip (S-), GeoCLIP, CSP, Engineer (Eng), Phi and Combined (Phi + Eng), Botswana \label{tab:lr_var}}
\begin{tblr}{
  cell{2}{3} = {c},
  cell{2}{4} = {c},
  cell{2}{5} = {c},
  cell{2}{6} = {c},
  cell{3}{3} = {c},
  cell{3}{4} = {c},
  cell{3}{5} = {c},
  cell{3}{6} = {c},
  cell{4}{3} = {c},
  cell{4}{4} = {c},
  cell{4}{5} = {c},
  cell{4}{6} = {c},
  cell{5}{3} = {c},
  cell{5}{4} = {c},
  cell{5}{5} = {c},
  cell{5}{6} = {c},
  cell{6}{3} = {c},
  cell{6}{4} = {c},
  cell{6}{5} = {c},
  cell{6}{6} = {c},
  cell{7}{3} = {c},
  cell{7}{4} = {c},
  cell{7}{5} = {c},
  cell{7}{6} = {c},
  cell{8}{3} = {c},
  cell{8}{4} = {c},
  cell{8}{5} = {c},
  cell{8}{6} = {c},
  cell{9}{3} = {c},
  cell{9}{4} = {c},
  cell{9}{5} = {c},
  cell{9}{6} = {c},
  cell{10}{3} = {c},
  cell{10}{4} = {c},
  cell{10}{5} = {c},
  cell{10}{6} = {c},
  cell{11}{3} = {c},
  cell{11}{4} = {c},
  cell{11}{5} = {c},
  cell{11}{6} = {c},
  cell{12}{3} = {c},
  cell{12}{4} = {c},
  cell{12}{5} = {c},
  cell{12}{6} = {c},
  vline{2} = {-}{},
  hline{1-2} = {-}{},
}
\textbf{Model} & \textbf{Features} & \textbf{Avg F1} & \textbf{F1 Variance} & \textbf{Avg Accuracy} & \textbf{Accuracy Variance} \\
               & S-R18-10          & 0.68            & 0.0000               & 0.52                  & 0.0000                     \\
               & S-R18-40          & 0.45            & 0.0000               & 0.54                  & 0.0000                     \\
               & S-R50-10          & 0.51            & 0.0000               & 0.4                   & 0.0000                     \\
               & S-R50-40          & 0.58            & 0.0000               & 0.55                  & 0.0000                     \\
               & S-V16-10          & 0.15            & 0.0000               & 0.52                  & 0.0000                     \\
\textbf{LR}    & S-V16-40          & 0.27            & 0.0000               & 0.47                  & 0.0000                     \\
               & GeoCLIP           & 0.7             & 0.0000               & 0.6                   & 0.0000                     \\
               & CSP               & 0.59            & 0.0000               & 0.51                  & 0.0000                     \\
               & Eng               & 0.67            & 0.0000               & 0.63                  & 0.0000                     \\
               & Phi               & 0.59            &     0.0000                 & 0.51                  &     0.0000                       \\
               & Phi + Eng         & 0.58            & 0.0000               & 0.55                  & 0.0000                     
\end{tblr}
\end{table}

\begin{table}[H]
\centering
\caption{Average F1 and Accuracy scores with variance over 5 runs for XGB Architecture for feature spaces SatClip (S-), GeoCLIP, CSP, Engineer (Eng), Phi and Combined (Phi + Eng), Botswana \label{tab:xgb_var}}
\begin{tblr}{
  cell{2}{3} = {c},
  cell{2}{4} = {c},
  cell{2}{5} = {c},
  cell{2}{6} = {c},
  cell{3}{3} = {c},
  cell{3}{4} = {c},
  cell{3}{5} = {c},
  cell{3}{6} = {c},
  cell{4}{3} = {c},
  cell{4}{4} = {c},
  cell{4}{5} = {c},
  cell{4}{6} = {c},
  cell{5}{3} = {c},
  cell{5}{4} = {c},
  cell{5}{5} = {c},
  cell{5}{6} = {c},
  cell{6}{3} = {c},
  cell{6}{4} = {c},
  cell{6}{5} = {c},
  cell{6}{6} = {c},
  cell{7}{3} = {c},
  cell{7}{4} = {c},
  cell{7}{5} = {c},
  cell{7}{6} = {c},
  cell{8}{3} = {c},
  cell{8}{4} = {c},
  cell{8}{5} = {c},
  cell{8}{6} = {c},
  cell{9}{3} = {c},
  cell{9}{4} = {c},
  cell{9}{5} = {c},
  cell{9}{6} = {c},
  cell{10}{3} = {c},
  cell{10}{4} = {c},
  cell{10}{5} = {c},
  cell{10}{6} = {c},
  cell{11}{3} = {c},
  cell{11}{4} = {c},
  cell{11}{5} = {c},
  cell{11}{6} = {c},
  cell{12}{3} = {c},
  cell{12}{4} = {c},
  cell{12}{5} = {c},
  cell{12}{6} = {c},
  vline{2} = {-}{},
  hline{1-2} = {-}{},
}
\textbf{Model} & \textbf{Features} & \textbf{Avg F1} & \textbf{F1 Variance} & \textbf{Avg Accuracy} & \textbf{Accuracy Variance} \\
               & S-R18-10          & 0.54            & 0.0000               & 0.6                   & 0.0000                     \\
               & S-R18-40          & 0.59            & 0.0000               & 0.62                  & 0.0000                     \\
               & S-R50-10          & 0.45            & 0.0008               & 0.53                  & 0.0005                     \\
               & S-R50-40          & 0.59            & 0.0002               & 0.6                   & 0.0002                     \\
               & S-V16-10          & 0.41            & 0.0000               & 0.53                  & 0.0000                     \\
\textbf{XGB}    & S-V16-40          & 0.4             & 0.0021               & 0.53                  & 0.0005                     \\
               & GeoCLIP           & 0.63            & 0.0005               & 0.6                   & 0.0011                     \\
               & CSP               & 0.52            & 0.0000               & 0.55                  & 0.0000                     \\
               & Eng               & 0.67            & 0.0023               & 0.64                  & 0.0014                     \\
               & Phi               & 0.52            &        0.0000              & 0.55                  &    0.000                        \\
               & Phi + Eng         & 0.68            & 0.0008               & 0.63                  & 0.0002                     
\end{tblr}
\end{table}

\begin{table}[H]
\centering
\caption{Average F1 and Accuracy scores with variance over 5 runs for RF Architecture for feature spaces SatClip (S-), GeoCLIP, CSP, Engineer (Eng), Phi and Combined (Phi + Eng), Rwanda \label{tab:rf_rwa_var}}
\begin{tblr}{
  cell{2}{3} = {c},
  cell{2}{4} = {c},
  cell{2}{5} = {c},
  cell{2}{6} = {c},
  cell{3}{3} = {c},
  cell{3}{4} = {c},
  cell{3}{5} = {c},
  cell{3}{6} = {c},
  cell{4}{3} = {c},
  cell{4}{4} = {c},
  cell{4}{5} = {c},
  cell{4}{6} = {c},
  cell{5}{3} = {c},
  cell{5}{4} = {c},
  cell{5}{5} = {c},
  cell{5}{6} = {c},
  cell{6}{3} = {c},
  cell{6}{4} = {c},
  cell{6}{5} = {c},
  cell{6}{6} = {c},
  cell{7}{3} = {c},
  cell{7}{4} = {c},
  cell{7}{5} = {c},
  cell{7}{6} = {c},
  cell{8}{3} = {c},
  cell{8}{4} = {c},
  cell{8}{5} = {c},
  cell{8}{6} = {c},
  cell{9}{3} = {c},
  cell{9}{4} = {c},
  cell{9}{5} = {c},
  cell{9}{6} = {c},
  cell{10}{3} = {c},
  cell{10}{4} = {c},
  cell{10}{5} = {c},
  cell{10}{6} = {c},
  cell{11}{3} = {c},
  cell{11}{4} = {c},
  cell{11}{5} = {c},
  cell{11}{6} = {c},
  cell{12}{3} = {c},
  cell{12}{4} = {c},
  cell{12}{5} = {c},
  cell{12}{6} = {c},
  vline{2} = {-}{},
  hline{1-2} = {-}{},
}
\textbf{Model} & \textbf{Features} & \textbf{Avg F1} & \textbf{F1 Variance} & \textbf{Avg Accuracy} & \textbf{Accuracy Variance} \\
               & S-R18-10          & 0.68            & 0.0001               & 0.63                  & 0.0001                     \\
               & S-R18-40          & 0.67            & 0.0000               & 0.62                  & 0.0000                     \\
               & S-R50-10          & 0.68            & 0.0001               & 0.63                  & 0.0000                     \\
               & S-R50-40          & 0.68            & 0.0000               & 0.63                  & 0.0000                     \\
               & S-V16-10          & 0.68            & 0.0002               & 0.63                  & 0.0003                     \\
\textbf{RF}    & S-V16-40          & 0.68            & 0.0001               & 0.62                  & 0.0001                     \\
               & GeoCLIP           & 0.69            & 0.0001               & 0.62                  & 0.0001                     \\
               & CSP               & 0.65            & 0.0000               & 0.6                   & 0.0000                     \\
               & Eng               & 0.71            & 0.0000               & 0.63                  & 0.0001                     \\
               & Phi               & 0.69            & 0.0001               & 0.55                  & 0.0003                     \\
               & Phi + Eng         & 0.69            & 0.0000               & 0.56                  & 0.0000                     
\end{tblr}
\end{table}

\begin{table}[H]
\centering
\caption{Average F1 and Accuracy scores with variance over 5 runs for MLP Architecture for feature spaces SatClip (S-), GeoCLIP, CSP, Engineer (Eng), Phi and Combined (Phi + Eng), Rwanda \label{mlp_rwa_var}}
\begin{tblr}{
  cell{2}{3} = {c},
  cell{2}{4} = {c},
  cell{2}{5} = {c},
  cell{2}{6} = {c},
  cell{3}{3} = {c},
  cell{3}{4} = {c},
  cell{3}{5} = {c},
  cell{3}{6} = {c},
  cell{4}{3} = {c},
  cell{4}{4} = {c},
  cell{4}{5} = {c},
  cell{4}{6} = {c},
  cell{5}{3} = {c},
  cell{5}{4} = {c},
  cell{5}{5} = {c},
  cell{5}{6} = {c},
  cell{6}{3} = {c},
  cell{6}{4} = {c},
  cell{6}{5} = {c},
  cell{6}{6} = {c},
  cell{7}{3} = {c},
  cell{7}{4} = {c},
  cell{7}{5} = {c},
  cell{7}{6} = {c},
  cell{8}{3} = {c},
  cell{8}{4} = {c},
  cell{8}{5} = {c},
  cell{8}{6} = {c},
  cell{9}{3} = {c},
  cell{9}{4} = {c},
  cell{9}{5} = {c},
  cell{9}{6} = {c},
  cell{10}{3} = {c},
  cell{10}{4} = {c},
  cell{10}{5} = {c},
  cell{10}{6} = {c},
  cell{11}{3} = {c},
  cell{11}{4} = {c},
  cell{11}{5} = {c},
  cell{11}{6} = {c},
  cell{12}{3} = {c},
  cell{12}{4} = {c},
  cell{12}{5} = {c},
  cell{12}{6} = {c},
  vline{2} = {-}{},
  hline{1-2} = {-}{},
}
\textbf{Model} & \textbf{Features} & \textbf{Avg F1} & \textbf{F1 Variance} & \textbf{Avg Accuracy} & \textbf{Accuracy Variance} \\
               & S-R18-10          & 0.66            & 0.0014               & 0.6                   & 0.0004                     \\
               & S-R18-40          & 0.67            & 0.0007               & 0.6                   & 0.0007                     \\
               & S-R50-10          & 0.62            & 0.0054               & 0.62                  & 0.0020                     \\
               & S-R50-40          & 0.63            & 0.0052               & 0.61                  & 0.0005                     \\
               & S-V16-10          & 0.61            & 0.0093               & 0.6                   & 0.0008                     \\
\textbf{MLP}   & S-V16-40          & 0.59            & 0.0108               & 0.57                  & 0.0018                     \\
               & GeoCLIP           & 0.65            & 0.0010               & 0.6                   & 0.0010                     \\
               & CSP               & 0.63            & 0.0178               & 0.58                  & 0.0018                     \\
               & Eng               & 0.67            & 0.0007               & 0.59                  & 0.0010                     \\
               & Phi               & 0.62            & 0.0086               & 0.52                  & 0.0003                     \\
               & Phi + Eng         & 0.69            & 0.0001               & 0.53                  & 0.0001                     
\end{tblr}
\end{table}

\begin{table}[H]
\centering
\caption{Average F1 and Accuracy scores with variance over 5 runs for GB Architecture for feature spaces SatClip (S-), GeoCLIP, CSP, Engineer (Eng), Phi and Combined (Phi + Eng), Rwanda \label{gb_rwa_var}}
\begin{tblr}{
  cell{2}{3} = {c},
  cell{2}{4} = {c},
  cell{2}{5} = {c},
  cell{2}{6} = {c},
  cell{3}{3} = {c},
  cell{3}{4} = {c},
  cell{3}{5} = {c},
  cell{3}{6} = {c},
  cell{4}{3} = {c},
  cell{4}{4} = {c},
  cell{4}{5} = {c},
  cell{4}{6} = {c},
  cell{5}{3} = {c},
  cell{5}{4} = {c},
  cell{5}{5} = {c},
  cell{5}{6} = {c},
  cell{6}{3} = {c},
  cell{6}{4} = {c},
  cell{6}{5} = {c},
  cell{6}{6} = {c},
  cell{7}{3} = {c},
  cell{7}{4} = {c},
  cell{7}{5} = {c},
  cell{7}{6} = {c},
  cell{8}{3} = {c},
  cell{8}{4} = {c},
  cell{8}{5} = {c},
  cell{8}{6} = {c},
  cell{9}{3} = {c},
  cell{9}{4} = {c},
  cell{9}{5} = {c},
  cell{9}{6} = {c},
  cell{10}{3} = {c},
  cell{10}{4} = {c},
  cell{10}{5} = {c},
  cell{10}{6} = {c},
  cell{11}{3} = {c},
  cell{11}{4} = {c},
  cell{11}{5} = {c},
  cell{11}{6} = {c},
  cell{12}{3} = {c},
  cell{12}{4} = {c},
  cell{12}{5} = {c},
  cell{12}{6} = {c},
  vline{2} = {-}{},
  hline{1-2} = {-}{},
}
\textbf{Model} & \textbf{Features} & \textbf{Avg F1} & \textbf{F1 Variance} & \textbf{Avg Accuracy} & \textbf{Accuracy Variance} \\
               & S-R18-10          & 0.7             & 0.0002               & 0.63                  & 0.0001                     \\
               & S-R18-40          & 0.68            & 0.0018               & 0.62                  & 0.0016                     \\
               & S-R50-10          & 0.71            & 0.0026               & 0.65                  & 0.0020                     \\
               & S-R50-40          & 0.68            & 0.0011               & 0.63                  & 0.0006                     \\
               & S-V16-10          & 0.71            & 0.0006               & 0.65                  & 0.0006                     \\
\textbf{GB}    & S-V16-40          & 0.68            & 0.0029               & 0.62                  & 0.0014                     \\
               & GeoCLIP           & 0.69            & 0.0001               & 0.63                  & 0.0002                     \\
               & CSP               & 0.66            & 0.0017               & 0.59                  & 0.0010                     \\
               & Eng               & 0.7             & 0.0001               & 0.61                  & 0.0002                     \\
               & Phi               & 0.63            & 0.0007               & 0.53                  & 0.0002                     \\
               & Phi + Eng         & 0.63            & 0.0011               & 0.56                  & 0.0001                     
\end{tblr}
\end{table}

\begin{table}[H]
\centering
\caption{Average F1 and Accuracy scores with variance over 5 runs for SVM Architecture for feature spaces SatClip (S-), GeoCLIP, CSP, Engineer (Eng), Phi and Combined (Phi + Eng), Rwanda \label{svm_rwa_var}}
\begin{tblr}{
  cell{2}{3} = {c},
  cell{2}{4} = {c},
  cell{2}{5} = {c},
  cell{2}{6} = {c},
  cell{3}{3} = {c},
  cell{3}{4} = {c},
  cell{3}{5} = {c},
  cell{3}{6} = {c},
  cell{4}{3} = {c},
  cell{4}{4} = {c},
  cell{4}{5} = {c},
  cell{4}{6} = {c},
  cell{5}{3} = {c},
  cell{5}{4} = {c},
  cell{5}{5} = {c},
  cell{5}{6} = {c},
  cell{6}{3} = {c},
  cell{6}{4} = {c},
  cell{6}{5} = {c},
  cell{6}{6} = {c},
  cell{7}{3} = {c},
  cell{7}{4} = {c},
  cell{7}{5} = {c},
  cell{7}{6} = {c},
  cell{8}{3} = {c},
  cell{8}{4} = {c},
  cell{8}{5} = {c},
  cell{8}{6} = {c},
  cell{9}{3} = {c},
  cell{9}{4} = {c},
  cell{9}{5} = {c},
  cell{9}{6} = {c},
  cell{10}{3} = {c},
  cell{10}{4} = {c},
  cell{10}{5} = {c},
  cell{10}{6} = {c},
  cell{11}{3} = {c},
  cell{11}{4} = {c},
  cell{11}{5} = {c},
  cell{11}{6} = {c},
  cell{12}{3} = {c},
  cell{12}{4} = {c},
  cell{12}{5} = {c},
  cell{12}{6} = {c},
  vline{2} = {-}{},
  hline{1-2} = {-}{},
}
\textbf{Model} & \textbf{Features} & \textbf{Avg F1} & \textbf{F1 Variance} & \textbf{Avg Accuracy} & \textbf{Accuracy Variance} \\
               & S-R18-10          & 0.71            & 0.0000               & 0.64                  & 0.0000                     \\
               & S-R18-40          & 0.7             & 0.0000               & 0.54                  & 0.0000                     \\
               & S-R50-10          & 0.7             & 0.0000               & 0.54                  & 0.0000                     \\
               & S-R50-40          & 0.7             & 0.0000               & 0.54                  & 0.0000                     \\
               & S-V16-10          & 0.69            & 0.0000               & 0.59                  & 0.0000                     \\
\textbf{SVM}   & S-V16-40          & 0.7             & 0.0000               & 0.54                  & 0.0000                     \\
               & GeoCLIP           & 0.71            & 0.0000               & 0.64                  & 0.0000                     \\
               & CSP               & 0.72            & 0.0000               & 0.61                  & 0.0000                     \\
               & Eng               & 0.72            & 0.0000               & 0.62                  & 0.0000                     \\
               & Phi               & 0.7             & 0.0000               & 0.55                  & 0.0000                     \\
               & Phi + Eng         & 0.7             & 0.0000               & 0.56                  & 0.0000                     
\end{tblr}
\end{table}

\begin{table}[H]
\centering
\caption{Average F1 and Accuracy scores with variance over 5 runs for LR Architecture for feature spaces SatClip (S-), GeoCLIP, CSP, Engineer (Eng), Phi and Combined (Phi + Eng), Rwanda \label{lr_rwa_var}}
\begin{tblr}{
  cell{2}{3} = {c},
  cell{2}{4} = {c},
  cell{2}{5} = {c},
  cell{2}{6} = {c},
  cell{3}{3} = {c},
  cell{3}{4} = {c},
  cell{3}{5} = {c},
  cell{3}{6} = {c},
  cell{4}{3} = {c},
  cell{4}{4} = {c},
  cell{4}{5} = {c},
  cell{4}{6} = {c},
  cell{5}{3} = {c},
  cell{5}{4} = {c},
  cell{5}{5} = {c},
  cell{5}{6} = {c},
  cell{6}{3} = {c},
  cell{6}{4} = {c},
  cell{6}{5} = {c},
  cell{6}{6} = {c},
  cell{7}{3} = {c},
  cell{7}{4} = {c},
  cell{7}{5} = {c},
  cell{7}{6} = {c},
  cell{8}{3} = {c},
  cell{8}{4} = {c},
  cell{8}{5} = {c},
  cell{8}{6} = {c},
  cell{9}{3} = {c},
  cell{9}{4} = {c},
  cell{9}{5} = {c},
  cell{9}{6} = {c},
  cell{10}{3} = {c},
  cell{10}{4} = {c},
  cell{10}{5} = {c},
  cell{10}{6} = {c},
  cell{11}{3} = {c},
  cell{11}{4} = {c},
  cell{11}{5} = {c},
  cell{11}{6} = {c},
  cell{12}{3} = {c},
  cell{12}{4} = {c},
  cell{12}{5} = {c},
  cell{12}{6} = {c},
  vline{2} = {-}{},
  hline{1-2} = {-}{},
}
\textbf{Model} & \textbf{Features} & \textbf{Avg F1} & \textbf{F1 Variance} & \textbf{Avg Accuracy} & \textbf{Accuracy Variance} \\
               & S-R18-10          & 0.67            & 0.0000               & 0.59                  & 0.0000                     \\
               & S-R18-40          & 0.67            & 0.0000               & 0.6                   & 0.0000                     \\
               & S-R50-10          & 0.67            & 0.0000               & 0.61                  & 0.0000                     \\
               & S-R50-40          & 0.67            & 0.0000               & 0.59                  & 0.0000                     \\
               & S-V16-10          & 0.67            & 0.0000               & 0.61                  & 0.0000                     \\
\textbf{LR}    & S-V16-40          & 0.67            & 0.0000               & 0.59                  & 0.0000                     \\
               & GeoCLIP           & 0.69            & 0.0000               & 0.64                  & 0.0000                     \\
               & CSP               & 0.7             & 0.0000               & 0.6                   & 0.0000                     \\
               & Eng               & 0.7             & 0.0000               & 0.59                  & 0.0000                     \\
               & Phi               & 0.68            & 0.0000               & 0.55                  & 0.0000                     \\
               & Phi + Eng         & 0.68            & 0.0000               & 0.55                  & 0.0000                     
\end{tblr}
\end{table}

\begin{table}[H]
\centering
\caption{Average F1 and Accuracy scores with variance over 5 runs for XGB Architecture for feature spaces SatClip (S-), GeoCLIP, CSP, Engineer (Eng), Phi and Combined (Phi + Eng), Rwanda \label{tab:xgb_rwa_var}}
\begin{tblr}{
  cell{2}{3} = {c},
  cell{2}{4} = {c},
  cell{2}{5} = {c},
  cell{2}{6} = {c},
  cell{3}{3} = {c},
  cell{3}{4} = {c},
  cell{3}{5} = {c},
  cell{3}{6} = {c},
  cell{4}{3} = {c},
  cell{4}{4} = {c},
  cell{4}{5} = {c},
  cell{4}{6} = {c},
  cell{5}{3} = {c},
  cell{5}{4} = {c},
  cell{5}{5} = {c},
  cell{5}{6} = {c},
  cell{6}{3} = {c},
  cell{6}{4} = {c},
  cell{6}{5} = {c},
  cell{6}{6} = {c},
  cell{7}{3} = {c},
  cell{7}{4} = {c},
  cell{7}{5} = {c},
  cell{7}{6} = {c},
  cell{8}{3} = {c},
  cell{8}{4} = {c},
  cell{8}{5} = {c},
  cell{8}{6} = {c},
  cell{9}{3} = {c},
  cell{9}{4} = {c},
  cell{9}{5} = {c},
  cell{9}{6} = {c},
  cell{10}{3} = {c},
  cell{10}{4} = {c},
  cell{10}{5} = {c},
  cell{10}{6} = {c},
  cell{11}{3} = {c},
  cell{11}{4} = {c},
  cell{11}{5} = {c},
  cell{11}{6} = {c},
  cell{12}{3} = {c},
  cell{12}{4} = {c},
  cell{12}{5} = {c},
  cell{12}{6} = {c},
  vline{2} = {-}{},
  hline{1-2} = {-}{},
}
\textbf{Model} & \textbf{Features} & \textbf{Avg F1} & \textbf{F1 Variance} & \textbf{Avg Accuracy} & \textbf{Accuracy Variance} \\
               & S-R18-10          & 0.71            & 0.0002               & 0.65                  & 0.0002                     \\
               & S-R18-40          & 0.72            & 0.0000               & 0.65                  & 0.0000                     \\
               & S-R50-10          & 0.69            & 0.0005               & 0.63                  & 0.0003                     \\
               & S-R50-40          & 0.71            & 0.0008               & 0.63                  & 0.0006                     \\
               & S-V16-10          & 0.69            & 0.0000               & 0.63                  & 0.0000                     \\
\textbf{XGB}   & S-V16-40          & 0.71            & 0.0001               & 0.64                  & 0.0002                     \\
               & GeoCLIP           & 0.7             & 0.0002               & 0.63                  & 0.0003                     \\
               & CSP               & 0.69            & 0.0000               & 0.61                  & 0.0000                     \\
               & Eng               & 0.67            & 0.0005               & 0.61                  & 0.0006                     \\
               & Phi               & 0.66            & 0.0007               & 0.54                  & 0.0002                     \\
               & Phi + Eng         & 0.67            & 0.0007               & 0.59                  & 0.0002                     
\end{tblr}
\end{table}

\end{document}